\documentclass[10pt,journal,compsoc]{IEEEtran}
\usepackage{algorithmic}
\usepackage{algorithm}
\usepackage{array}
\usepackage{textcomp}
\usepackage{stfloats}
\usepackage{xurl}
\usepackage{verbatim}
\usepackage{graphicx}
\usepackage{multirow}
\usepackage{subfigure}
\usepackage{caption}
\usepackage{bbm}
\usepackage{soul}
\usepackage{todonotes}
\usepackage{xcolor,colortbl}
\definecolor{LightGreen}{rgb}{0.8,1,0.89}
\definecolor{LightRed}{rgb}{1.0,0.8,0.7}
\definecolor{LightCyan}{rgb}{0.88,1,1}
\usepackage{times}
\usepackage{latexsym}
\usepackage{blindtext}
\usepackage{amsmath,amssymb}
\usepackage{arydshln}
\usepackage{booktabs}
\usepackage{color}
\usepackage{soul}
\usepackage{wrapfig}
\usepackage{subcaption}
\usepackage[hidelinks]{hyperref}

\usepackage[inline]{enumitem}
\usepackage{amsmath, amsfonts, mathtools}
\usepackage{tikz}

\newcommand{\ExternalLink}{%
    \tikz[x=1.2ex, y=1.2ex, baseline=-0.05ex]{%
        \begin{scope}[x=1ex, y=1ex]
            \clip (-0.1,-0.1) 
                --++ (-0, 1.2) 
                --++ (0.6, 0) 
                --++ (0, -0.6) 
                --++ (0.6, 0) 
                --++ (0, -1);
            \path[draw, 
                line width = 0.5, 
                rounded corners=0.5] 
                (0,0) rectangle (1,1);
        \end{scope}
        \path[draw, line width = 0.5] (0.5, 0.5) 
            -- (1, 1);
        \path[draw, line width = 0.5] (0.6, 1) 
            -- (1, 1) -- (1, 0.6);
        }
    }

\newcommand{\dataset}{\texttt{MOOD}}
\newcommand{\harmeme}{\texttt{HarMeme}}
\newcommand{\dank}{\texttt{Dank Memes}}
\newcommand{\memotion}{\texttt{Memotion}}
\newcommand{\sent}{\texttt{SENT}}
\newcommand{\emot}{\texttt{EMOT}}
\newcommand{\emotq}{\texttt{EMOT-Q}}
\newcommand{\proposed}{\texttt{ALFRED}}

\ifCLASSOPTIONcompsoc
  \usepackage[nocompress]{cite}
\else
  \usepackage{cite}
\fi
\ifCLASSINFOpdf
\else
\fi
\begin{document}
%
\title{Emotion-Aware Multimodal Fusion for\\ Meme Emotion Detection}
%
%
%
%

\author{Shivam Sharma$^{1,4}$, Ramaneswaran S$^{3}$, Md. Shad Akhtar$^2$ and Tanmoy Chakraborty$^1$\\
  $^1$\textit{IIT Delhi, India}; $^2$\textit{IIIT Delhi, India}; $^3$\textit{VIT, Vellore, India}; $^4$\textit{Wipro AI Labs, India} \\
  \small\texttt{\{shivam.sharma,tanchak\}@ee.iitd.ac.in},  \small\texttt{shad.akhtar@iiitd.ac.in}, \small\texttt{s.ramaneswaran2000@gmail.com}
}

\markboth{IEEE TRANSACTIONS ON AFFECTIVE COMPUTING}%
{Shell \MakeLowercase{\textit{et al.}}: Bare Demo of IEEEtran.cls for Computer Society Journals}
%



\IEEEtitleabstractindextext{%
\begin{abstract}
The ever-evolving social media discourse has witnessed an overwhelming use of memes to express opinions or dissent. Besides being misused for spreading malcontent, they are mined by corporations and political parties to glean the public's opinion. Therefore, memes predominantly offer affect-enriched insights towards ascertaining the societal psyche. However, the current approaches are yet to model the affective dimensions expressed in memes effectively. They rely extensively on large multimodal datasets for pre-training and do not generalize well due to constrained visual-linguistic grounding. In this paper, we introduce \dataset\ (Meme emOtiOns Dataset), which embodies six basic emotions. We then present \proposed\ (emotion-Aware muLtimodal Fusion foR Emotion Detection), a novel multimodal neural framework that (i) explicitly models emotion-enriched visual cues, and (ii) employs an efficient cross-modal fusion via a gating mechanism. Our investigation establishes \proposed's superiority over existing baselines by 4.94\% F1. Additionally, \proposed\ competes strongly with previous best approaches on the challenging \memotion\ task. We then discuss \proposed's\ domain-agnostic generalizability by demonstrating its dominance on two recently-released datasets -- \harmeme\ and \dank, over other baselines. Further, we analyze \proposed's interpretability using attention maps. Finally, we highlight the inherent challenges posed by the complex interplay of disparate modality-specific cues toward meme analysis.
\end{abstract}

}

\maketitle

\IEEEdisplaynontitleabstractindextext

%
\IEEEpeerreviewmaketitle

\IEEEraisesectionheading{\section{Introduction}\label{sec:introduction}}

%
%
%
%
\IEEEPARstart{M}{emes} on social media represent a new digital artifact genre that has become ubiquitous with time. Internet memes contain a short piece of text embedded over an image, often expressing information sarcastically or humorously, but sometimes in illicit ways. 
Internet memes also offer rich potential to understand or sway the sentiment and opinion of communities on social media, essentially facilitating a systematic study of their affective characteristics. 

The increase in the amount of multimodal content being disseminated over the web has spurred innovation in allied areas involving multimodality in general; however, fewer efforts have been made toward analyzing memes, especially to the extent its inherent complexity solicits. Memes, having multimodal constructs, require a joint interpretation of both the embedded text and the visuals to assimilate the intended meaning comprehensively. Although existing approaches perform much better on conventional multimodal tasks, particularly the ones involving visual-linguistic grounding \cite{trends2021}, they are yet to deliver for other scenarios like memes \cite{sharma2020semeval2020,DBLP:conf/aaai/SharmaASNA023,DBLP:conf/acl/SharmaSAA023,DBLP:conf/ijcnlp/SharmaSAC22,DBLP:conf/emnlp/PramanickSDAN021}.

\begin{figure}
    \centering
    \includegraphics[width=0.95\columnwidth]{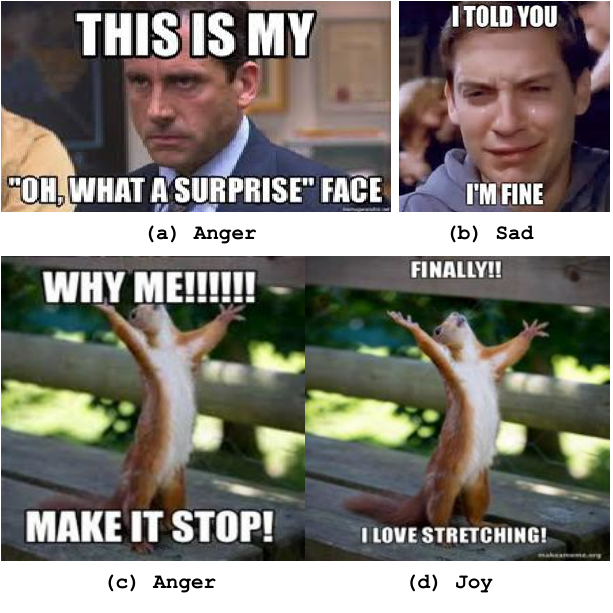}
    \caption{Example of memes: (a,b) text modality is misleading; (c,d) same image but emotion being differentiated by the text.}
    \label{fig:meme_challenge}
    \vspace{-5mm}
\end{figure}

One of the critical challenges within the studies related to social media content analysis is the \textit{subjective perception problem} \cite{subjective_perception_ijcai}, which leads to ambiguous data labeling. Consequently, several recent participatory efforts involving the detection of emotion from memes consider the categories that capture higher levels of affective abstraction like \textit{humor, sarcasm, offense, and motivation} \cite{sharma2020semeval2020,memotion2}. Although these efforts effectively capture affective phenomena from memes that pertain to macro societal aspects like political, socio-cultural, demographics, etc., they constrain fine-grained analysis of meme emotions. Efforts are needed towards studies that target more fine-grained analysis of user behavior, decisions, and perceptions \cite{socialmediamarketing}, encompassing a broader spectrum of emotions that help address the multimodal affective characterization at the fundamental level. 

Additionally, the inherent multimodality involving image-text configurations presents an additional challenge in detecting emotions due to the complexity posed by the implicit background context abstracted by \textit{memetic visuals}. For instance, memes shown in Figs. \ref{fig:meme_challenge}(a) and \ref{fig:meme_challenge}(b) depict faces with expressions conveying \textit{anger} and \textit{sadness}, respectively; whereas the corresponding embedded text suggests something different, thereby vesting the required onus of meme emotion detection, primarily upon the visual cues. Also, examples shown in Figs. \ref{fig:meme_challenge}(c) and \ref{fig:meme_challenge}(d) carry the same background image, but their corresponding embedded texts convey contrasting emotions -- \textit{anger} and \textit{joy}, respectively, indicating triviality of the visual cues present. These aspects pose challenges like visual-linguistic dissociation and cross-modal noise, and induce reasoning complexity by having dependencies on implicit contextual cues, effectively inhibiting the overall progress towards addressing such non-trivial tasks.

In this work, we propose a novel task of classifying the emotions expressed within memes amongst \textit{six} basic Ekman \cite{basic_emotions_ekman} emotions. To this end, we also introduce a manually curated large-scale multimodal dataset, \dataset\ (Meme emOtiOns Dataset). \dataset\ constitutes real multimodal memes (images with overlaid text) expressing various emotions that are discretely mapped to the six fundamental Ekman \cite{basic_emotions_ekman} emotions -- \textit{fear, anger, joy, sadness, surprise,} and {\em disgust} via manual annotation. We benchmark this dataset against several unimodal and multimodal systems emulating competitive baselines for meme emotion detection and report class-wise, along with macro-averaged performances. Further, we investigate the design of an effective approach to detect emotions from memes and propose \proposed, a multimodal approach that employs systematic modality-specific interactions via gating. \proposed\ constitutes (a) gated multimodal fusion (GMF) towards explicitly incorporating emotion-enriched visual features, followed by (b) gated cross-attention (GCA) to fuse emotion-enriched image and text representations, conditioned upon the visual cues learned. We observe significant gains by \proposed\ over other strong baselines on the \dataset\ dataset, along with competitive performance on \memotion\ tasks \cite{sharma2020semeval2020}, followed by distinct indications of \proposed's strong generalizability over other related tasks such as  \harmeme\ \cite{harmeme} and \dank \cite{dank_evalita}.  

We also examine the interpretability of the predictions by analyzing visual attention maps corresponding to the encoding mechanisms that \proposed\ employs and highlight both affective affinity and limitations exhibited therein. Besides discussing and analyzing the overall and emotion-specific performances of various systems and \proposed\ in detail, we delineate the contributory aspects of \dataset\ and \proposed. Finally, we perform an extensive error analysis elucidating some imminent challenges and limitations like \textit{modality-specific obscurity} and \textit{thematic overlaps} rendered by the complex memetic dynamics.

Through this work, we intend to address the imperative necessity of characterizing multimodal content like memes and their fundamental affective spectrum by emphasizing their esoteric visual semiotics. In particular, we make the following contributions:

\begin{itemize}[leftmargin=1.5em]
  \item We introduce \dataset\ (Meme emOtiOns Dataset) that captures six basic \textit{Ekman's} emotions for memes.
  \item We propose \proposed\ (emotion-Aware muLtimodal Fusion foR Emotion Detection): a multimodal neural framework that uses affect-enriched features from memes and fuses them via a gated cross-attention mechanism.
  \item We benchmark the dataset via several unimodal and multimodal baselines and discuss their limitations. 
  \item Further, we empirically demonstrate the efficacy of \proposed\ over strong baselines on the \dataset\ and \memotion\ datasets.  
  \item Finally, we perform interpretability analysis and establish the generalizability of \proposed\ on meme datasets capturing distributions beyond six basic Ekman emotions. 
\end{itemize}

\noindent \textbf{Reproducibility:} The source codes and the sample dataset are uploaded at: {\footnotesize\texttt{\url{https://github.com/LCS2-IIITD/ALFRED_MemeEmotionDetection}}}.

\section{Related Work}

Several studies contributed to the understanding of detecting emotions from multimodal content, encompassing modalities like images and text, along with signals like audio, video, EEG, eye movement, etc. There have also been numerous recent efforts toward analyzing memes and detecting various allied harmful aspects. Since there have been limited exploratory studies on detecting emotions from memes, the field can benefit significantly from the findings of the aforementioned applications. We systematically review these areas to set the necessary background.  \\

\vspace{-1mm}

\noindent \textbf{Multimodal emotion detection:}
Emotion detection is a well-studied area explored for various modalities such as text, speech, and audio \cite{abdul-mageed-ungar-2017-emonet,affectnet}. Significant emphasis has been laid on multimodal emotion detection as well. Unlike traditional emotion detection tasks involving single modalities, multimodal approaches require a mechanism to effectively learn features from multiple correlated modalities. An initial effort in this domain \cite{poria_multimodal_emotion_2016} proposed multi-kernel learning based deep-CNN towards emotion and sentiment recognition on different multimodal datasets. This was followed by proffering a pooling-based fusion mechanism in \cite{duong2017multimodal} along with introducing a multimodal social media dataset (and metadata) from Reddit towards the domain of emotion classification. Efforts have been made toward multimodal feature characterization \cite{hu_tumblr2018}, wherein authors introduced a Tumblr-based multimodal dataset and demonstrated the efficacy of multimodal approaches when contrasted with their unimodal counterparts. Another notable finding \cite{correlated_attention_2018} explored an approach advocating coordinated representation learning for multimodal emotion recognition. The authors used a recurrent neural network to emulate correlated attention and calculate the correlation between EEG and eye movement signals. 

Recent efforts include a Transformer-based inter-modality attention mechanism \cite{mm_transformer_fusion_2020} with self-supervision \cite{transformer_fusion_2020}, while materializing the design for fusing features from different modalities for multimodal emotion recognition. While there has been significant progress from a computational standpoint toward emotion recognition, Mittal et al. \cite{mittal2020emoticon} attempted to investigate an approach based on Frege's Context principle, which provides different interpretations of context for emotion recognition. The authors studied different interpretations using modality-specific features, semantic content, and depth-map to develop their algorithm. Although these efforts pave the way for addressing critical challenges prevalent within the affect-oriented applications for multimodal content, there is still scope for further exploring multimodal content representing dynamic cross-modal semiotics, like memes. \\

\vspace{-1mm}
\noindent \textbf{Meme analysis:}
A significant influx of memes from online fringe communities, such as Gab, Reddit, and 4chan, to mainstream platforms, such as Twitter and Instagram, resulted in a massive epidemic of intended harm \cite{Zannettou2018}. This has imminently solicited addressing the prevailing challenges from the computational social science point of view. Towards this end, several datasets capturing offensiveness \cite{suryawanshi-etal-2020-multimodal}, hatefulness \cite{kiela2020hateful,gomez2019exploring}, and harmfulness \cite{pramanick-etal-2021-momenta-multimodal} in memes have been curated. Besides the tasks corresponding to these resources, there are a variety of other tasks, such as detecting sexism \cite{Drakett2018}, racism \cite{blmBecky2020}, and harmful propaganda \cite{nordinNeoNaziAltright2021} from memes, that have been explored from the perspective of critical discourse analysis. 

Participatory events like the Facebook Hateful Meme Challenge \cite{kiela2020hateful} and shared-task on detecting \textit{hero, villain} and \textit{victim} from memes \cite{sharma-etal-2022-findings} have laid a strong foundation for community-level initiatives for detecting hate speech \cite{masud2021hate,garg2023handling,masud2022proactively,kulkarni2023revisiting,DBLP:journals/corr/abs-2401-16727} and connotative role-labels in memes. As part of these challenges, several interesting approaches besides ensembling large language models, utilizing meta information, attentive interactions, and adaptive loss are attempted in the multimodal setting \cite{das2020detecting,sandulescu2020detecting,zhou2020multimodal,lippe2020multimodal}. Other notable insights from meme analyses suggest the utility of commonsense knowledge \cite{9582340}, web entities, racial aspects \cite{pramanick-etal-2021-momenta-multimodal,karkkainen2019fairface}, and other external cues for detecting offense, harm, and hate speech in memes.

Most of these efforts either address the detection tasks at various levels for harmfulness (see a recent survey \cite{Survey:2022:Harmful:Memes}) or design ensemble techniques lacking cost-optimality. However, they tend to ignore the primary spectrum of emotions that plays a crucial role in cascading any adverse effect over social media. The current study aims to address this fundamental aspect of detecting the basic emotions of memes.\\

\vspace{-1mm}
\noindent \textbf{Emotion detection from memes:}
Although several studies have analyzed emotions of social media content, fewer efforts have been made toward characterizing the emotions of Internet memes. French \cite{French2017ImagebasedMA_meme_sent_pred} studied the correlation between the semantics of a meme and the textual discussions in the thread related to a multimodal post. Their study indicates the effectiveness of memes as a sentiment predictor over social media. \memotion\ \cite{sharma2020semeval2020,memotion2}, a series of participatory shared tasks on meme emotion classification, initiated the task of detecting meme emotions at different levels of granularity. Their dataset was curated towards three sub-tasks, emulating various combinations of \textit{multi-class/label} formulations. The three sub-tasks were -- (a) {\em sentiment classification}: multi-class classification amongst positive and negative categories; (b) {\em emotion classification}: multi-class classification amongst categories \textit{sarcastic, humorous, offensive and motivational}; and (c) {\em quantification}: multi-class/label classification amongst emotion intensities, represented by \textit{slightly, mildly} and \textit{very} and across emotion categories. 
Participants of this task explored several unimodal and multimodal approaches. For unimodal feature extraction, participants used a variety of models such as BERT \cite{devlin2019BERT}, GloVe \cite{pennington-etal-2014-glove} for text modality and pre-trained image models such as EfficientNet \cite{tan2020efficientnet} and ResNet \cite{he2015resnet} for image modality. Singh et al. \cite{Singh2020LT3AS_MTL} and Vlad et al. \cite{vlad2020upb_mtl} used multi-task learning to jointly predict emotion and sentiment. While these solutions, as discussed in the Introduction section (c.f. Sec. \ref{sec:introduction}), address the detection of affect categories at a higher level of abstraction, they do not consider the multimodal emotion characterization from a fundamental perspective. This effectively renders the investigation of basic emotions from memes obscure. 
 
\section{Meme emOtiOns Dataset (\dataset)}
Memotion dataset \cite{sharma2020semeval2020,memotion2} includes affective categories like \textit{motivation, offense, sarcasm}, and {\em humor}, which represent high-level emotion abstraction within memes. Although these categories are critical for studying the imminent implications of memetic discourse over social media, they are insufficient for characterizing their impact on an individual's psyche. Therefore, as part of this work, we aim to set up a framework for addressing emotion recognition from memes w.r.t. the basic emotions. Ekman and  Cordaro \cite{basic_emotions_ekman} empirically suggested that human beings exhibit six basic emotions, namely {\em fear} (FER), {\em anger} (AGR), {\em joy} (JOY), {\em sadness} (SDN), {\em surprise} (SPR), and {\em disgust} (DGT). 
Besides constituting the primary spectrum for studying the human affective response, basic emotions decide the overall affective tone of memes towards analyzing their disseminative outcomes. To this end, we manually curate \dataset, a multimodal dataset for detecting basic emotions from memes. 

\begin{table}[t]
\centering
\caption{Summary of \dataset\ \& Extended AffectNet.}
\resizebox{\columnwidth}{!}{
\begin{tabular}{ccccccccc}
\toprule
Dataset & Split & \# memes & FER & AGR & JOY & SAD & SPR & DGT \\ \midrule

\multirow{4}{*}{\rotatebox{0}{\dataset}} & Train & 7004  & 612 & 1413 & 1920 & 1822 & 855  & 382 \\
& Val & 1500  & 131 & 292  & 394  & 416  & 168  & 99  \\ 
& Test & 1500  & 128 & 312  & 392  & 398  & 185  & 85  \\ \cmidrule{2-9}
& Total & 10004 & 871 & 2017 & 2706 & 2636 & 1208 & 566 \\ \midrule

\multirow{3}{*}{\rotatebox{0}{\begin{tabular}[c]{@{}c@{}}Ext.\\Affect \\Net\end{tabular}}} & Subset & 50389 & 6540 & 10000 & 10000 & 10000 & 10000 & 3849 \\
& Add-On         & 1447  & 162  & 198   & 400   & 472   & 169   & 46   \\ \cmidrule{2-9}
& Total   & 51836 & 6702 & 10198 & 10400 & 10472 & 10169 & 3895 \\ \bottomrule
\end{tabular}}
\label{tab:data-distribution}
\end{table}

\subsection{Dataset collection and de-duplication}
\label{subsec:data}
The memes were collected primarily from two sources -- Google image search\footnote{\href{https://www.google.com/imghp?hl=en}{Google Images \ExternalLink}}
and imgflip\footnote{\href{https://imgflip.com}{Imgflip \ExternalLink}}.
We used keywords like `happy memes', `depression memes', `sad cat memes', etc., to crawl a diverse set of memes, capturing the six basic Ekman emotions. Many duplicates in the collected set were removed using an off-the-shelf API called imagededup\footnote{\href{https://github.com/idealo/imagededup.git}{imagededup \ExternalLink}}, followed by manually filtering out the low-quality memes. We used a set of filtering criteria for memes to ensure that the memes collected were of high quality. A meme is discarded if any of the following criteria are met: 
\begin{enumerate*}
    \item[(1)] The resolution of the meme is bad, such that the meme image is unclear or the readability of the meme text is affected;
    \item[(2)] If the meme did not exhibit any of the 6 Ekman emotions;
    \item[(3)] If the meme had content that induces hate towards or is harmful to individuals or communities;
    \item[(4)] Contain any personal information of a user;
    \item[(5)] Contains text that is not in English or is code-switched.
\end{enumerate*}This resulted in a total of $10,004$ memes (c.f. Table \ref{tab:data-distribution}), that constituted our primary proposed meme emotion dataset \dataset. The dataset constitutes \textit{generic} memes corresponding to the six basic Ekman emotions, on general topics like \textit{birthdays, relationships, family, friends} (to name a few), with an appropriate mix of human subjects, pop-culture references, animated characters, and animals.\footnote{See Appendix \ref{app:subsec:themes} for more details on thematic-distribution in \dataset}.

There is a variation in the relative proportions of the memes collected for different emotion categories. Many memes were collected from Google image search results for categories \textit{anger, joy, sadness, and disgust.}
Interestingly, almost half of the \textit{fear and surprise} memes we collected are obtained from imgflip. This suggests the categorical diversity that the platform can provide. Also, very few memes for categories \textit{disgust and sadness} could be sourced from imgflip, suggesting the positive sentiment-based genre that dominates the platform. This distribution also reflects the realistic availability of such multimodal data over different platforms.

\dataset\ consists of a total of 2017, 2706, and 2636 memes from categories \textit{anger, joy} and \textit{sadness}, respectively, constituting the majority share within the dataset. These are followed by the 1208 from \textit{surprise}, 871 from \textit{fear} and the least from disgust, with 576 memes. The summary of \dataset\ can be observed from Table \ref{tab:data-distribution}.

\begin{figure}[t!]
    \centering
    \includegraphics[width=0.8\columnwidth]{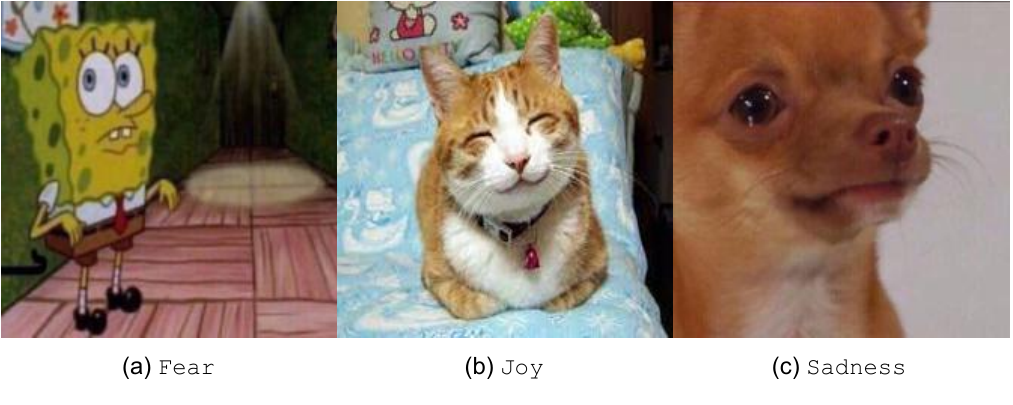}
    \caption{Examples of the images added to \textit{extend} AffectNet dataset, especially towards capturing the \textit{non-human} subjects within the meme visuals like, (a) \textit{fearful} `Spongebob', (b) \textit{joyous} cat and (c) \textit{sad} puppy, depicted in subfig. (a), (b) and (c), respectively.}
    \label{fig:example-reaction-images}
\end{figure}

Further, towards capturing the emotional signals expressed via visual modality, we leverage AffectNet \cite{affectnet}, a large-scale human facial expression dataset that consists of around $400K$ manually annotated facial expression images capturing \textit{neutral} and a total of 7 emotions, namely \textit{happy, sad, surprise, fear, disgust, anger} and \textit{contempt}. We randomly sample around $50K$ images corresponding to the Ekman emotion categories (excluding neutral and contempt from AffectNet) for our framework. AffectNet contains facial expressions only for \textit{humans}, rendering the animated and graphical emotion depiction and comprehension obscure, as required for our scenario. In order to robustly identify the emotions in realistic meme images within \dataset, especially the ones containing cartoons, animated figures, and animals, we needed to emphasize our multimodal framework's capability to handle them. To this end, we collected a dataset of $1,447$ images (c.f. Table \ref{tab:data-distribution}) with animated characters, cartoons, and animals and manually labeled them with Ekman emotion classes. We queried the web for using a simple formatted query string as `reactionary templates for' + $<$emotion category$>$ + $<$animal/cartoon$>$, and followed similar filtering/annotation process as for \dataset. 
These images represent various standard reactionary templates disseminated over social media. Fig. \ref{fig:example-reaction-images} depicts some examples of the images added to \textit{extend} the AffectNet dataset, especially towards capturing the \textit{non-human} subjects within the memes with query prefix as `reactionary template for', and suffixes as: \textit{fearful Spongebob}, \textit{joyous cat} and \textit{sad puppy}, depicted in Figs. \ref{fig:example-reaction-images}(a), \ref{fig:example-reaction-images}(b) and \ref{fig:example-reaction-images}(c), respectively. We call this extended set {\bf Ext. AffectNet}.

\begin{figure*}[t!]
    \centering
    \subfigure[Fear]{\includegraphics[width=0.16\textwidth]{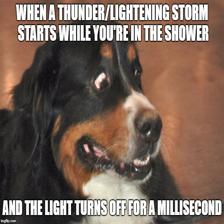}}
    \subfigure[Anger]{\includegraphics[width=0.16\textwidth]{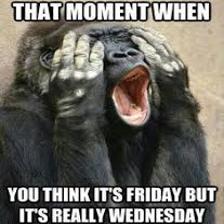}} 
    \subfigure[Joy]{\includegraphics[width=0.16\textwidth]{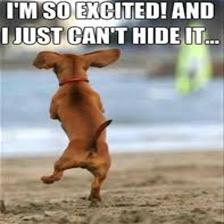}}
    \subfigure[Sadness]{\includegraphics[width=0.16\textwidth]{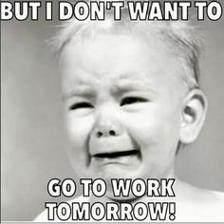}}
    \subfigure[Surprise]{\includegraphics[width=0.16\textwidth]{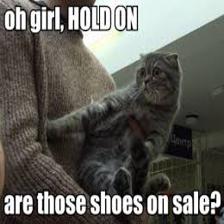}}    
    \subfigure[Disgust]{\includegraphics[width=0.16\textwidth]{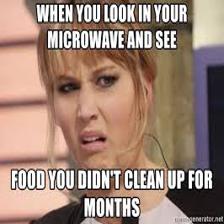}} 
    \caption{Examples depicting memes for six basic \textit{Ekman} emotions from our dataset, \dataset.}
    \label{fig:six_meme_eg}
\end{figure*}

\begin{figure}[!t]
\centering
\includegraphics[width=\columnwidth]{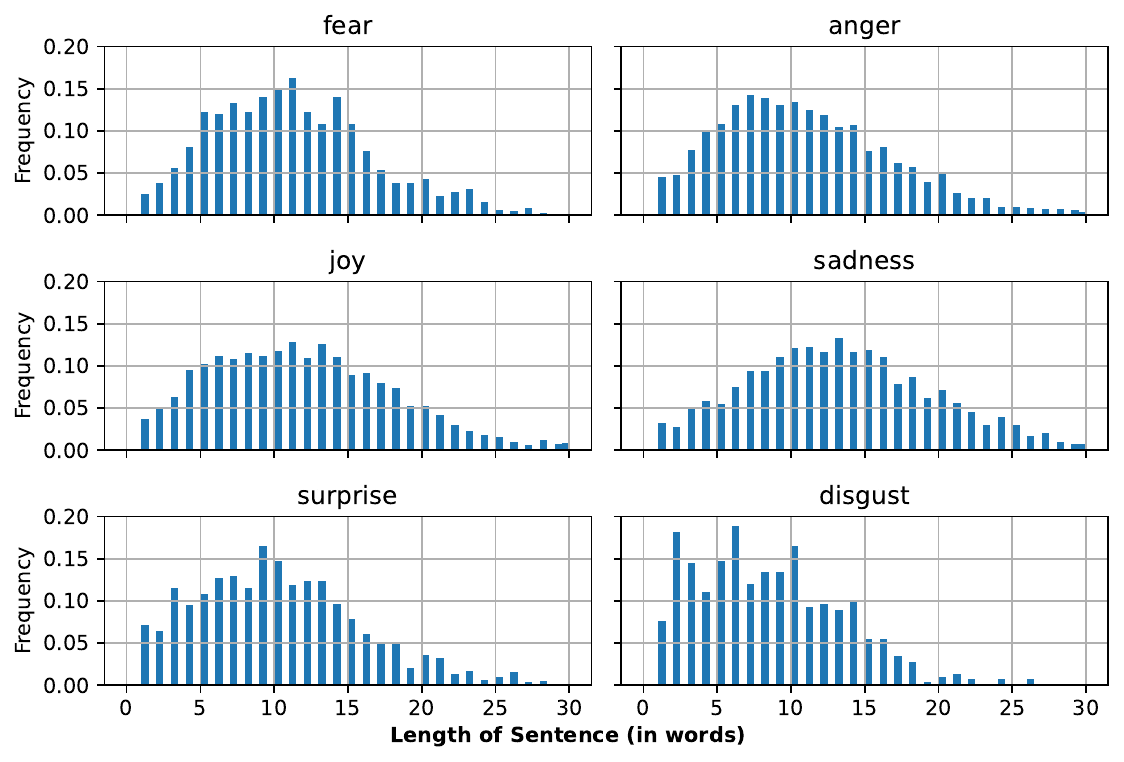}
\caption{Normalized histograms of meme text length per class.}
\label{fig:text_length_dist}
\vspace{-5mm}
\end{figure}

\subsection{Text Length Analysis}
\label{sec:text_length}
The meme text length analysis indicates the complexity that could be posed within a corpus. The more diverse the text-length distribution is for each category, the more difficult it could be to model the sequence and underlying association. 
It can be visualized from Fig. \ref{fig:text_length_dist} that the distributions for \textit{anger, sadness, and joy} are relatively more close to being \textit{normal}, also suggesting that there is a consistent pattern for the creation of content for memes from these categories. In contrast, \textit{disgust}, along with \textit{surprise and fear} (with slight variability), have relatively more variation regarding the text lengths used. This suggests the challenge it poses to the language models and the diversity with which such content is created online.

\begin{table}[t]
\centering
\caption{Prescribed guidelines for \dataset's annotation.}
\resizebox{\columnwidth}{!}{%
\begin{tabular}{cp{8cm}}
\toprule
& \multicolumn{1}{c}{\textbf{Annotation Guidelines}} \\
\midrule
1 & Emotion labeling should consider the meme author's perspective. \\
2 & Emotion labels should emphasize upon meme's content only. \\
3 & Emotion label should be one of the six basic Ekman emotions. \\
4 & Both textual and visual cues should be factored-in while annotating. \\
\multirow{2}{*}{5} & Pop-cultural or vernacular-oriented references must be queried from additional information sources wherever needed. \\
\multirow{2}{*}{6} & Sample leading to multiple possible interpretations, or lack of mapping to any of Ekman's six emotions, should be skipped. \\
\bottomrule
\end{tabular}%
}
\label{tab:annotations}
\vspace{-5mm}
\end{table}

\subsection{Annotation}
Two annotators annotated the dataset, while a consolidator oversaw the entire annotation exercise. One of the annotators is male, while the other female, and their ages range from 24-35 years. Moreover, both of them were professional lexicographers and social media savvy. On the other hand, the consolidator was an expert working in the fields of Computational Social Sciences and NLP. Before starting the annotation process, they were briefed on the task using detailed guidelines. They were requested to assess memes' textual and visual content towards the final adjudication of the meme emotion. In particular, they were asked to identify the emotion that the meme's author is trying to express via a \textit{multi-class} labeling setup. A tabulated list of prescribed guidelines adopted towards the annotation is shown in Table \ref{tab:annotations}. We conducted the annotation process in two stages -- a dry run and a final annotation stage. The Cohen's Kappa \cite{cohenkappa} was computed to assess the inter-annotator agreement  prior to the commencement of the final annotating process and was found to be nearly perfect with a score of $0.86$. 

\begin{figure}[t]
    \centering
    \includegraphics[width=0.8\columnwidth]{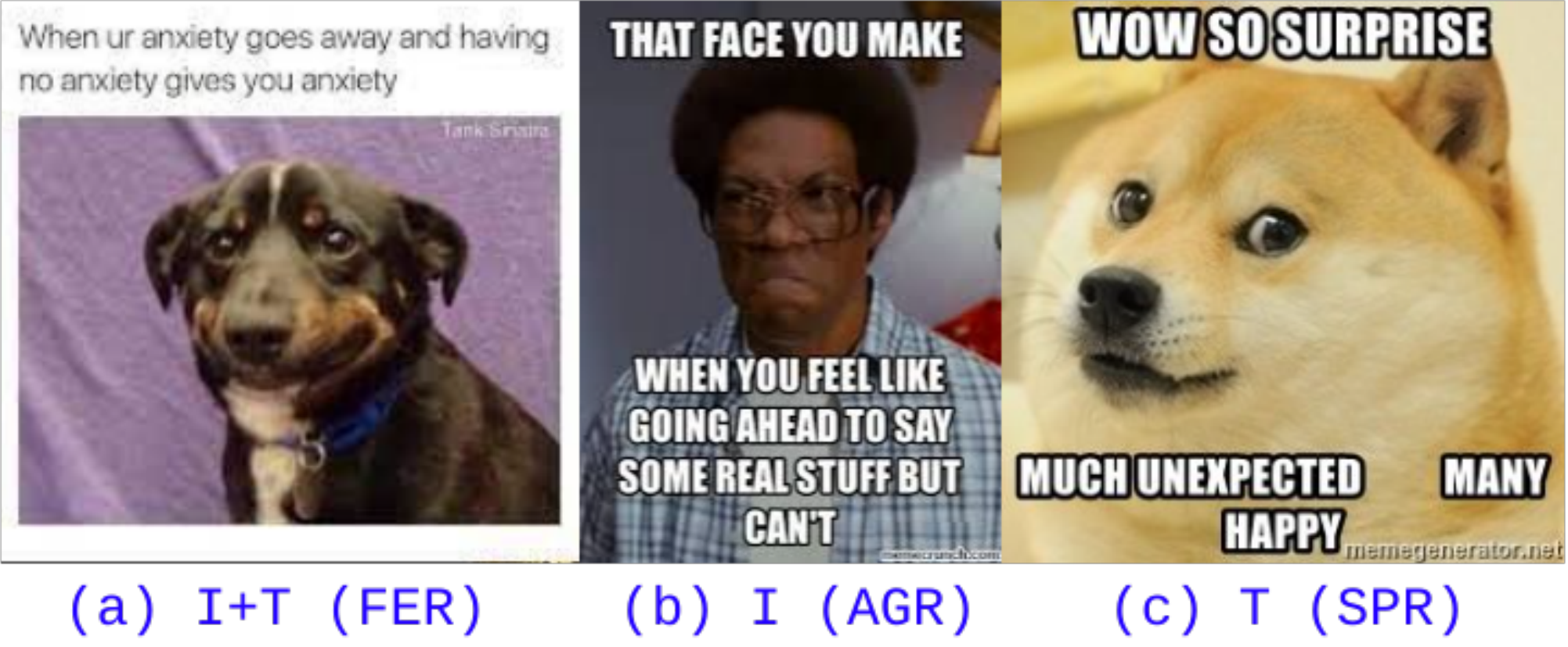}
    \caption{Example memes depicting modality-specific influence for emotion recognition. \textcolor{blue}{\texttt{I: Image, T: Text}.}}
    \label{fig:meme_eg}
\end{figure}

Based on the annotation, \dataset\ dataset can be exemplified via Fig. \ref{fig:six_meme_eg}, which depicts example memes for each of the six \textit{Ekman} emotions from the \dataset\ dataset. Although most memes in \dataset\ are designed by their authors to disseminate some form of humor via sarcasm, satire, or benign limericks, their primary objective is to resonate with the consumer's emotional appeal. Typically, this leads to the memes exhibiting a \textit{primary emotion} by design. Samples depicted in Figs. \ref{fig:six_meme_eg}(a)-\ref{fig:six_meme_eg}(f) are specially hand-picked to ensure a clear understanding of the annotation strategy. As can be observed from the samples, the primary emotions conveyed in the form of \textit{disgust, anger, surprise, fear, sadness} and \textit{joy} within the respective samples are expressed by both text and visual cues within the memes. The modality-specific expressivity varies extensively across \dataset, which constitutes the key multimodal challenge posed while performing analysis over realistic memes. The demonstration via Fig. \ref{fig:meme_eg} depicts the independent and joint influence of both image and text modalities via memes. Fig. \ref{fig:meme_eg}(a) depicts \textit{fear} through both text and image; Fig. \ref{fig:meme_eg}(b) shows \textit{anger} mainly via the facial expressions; and Fig. \ref{fig:meme_eg}(c) expresses \textit{surprise} only via text.

\begin{figure*}[t!]
    \centering
    \subfigure[Fear]{\includegraphics[width=0.16\textwidth]{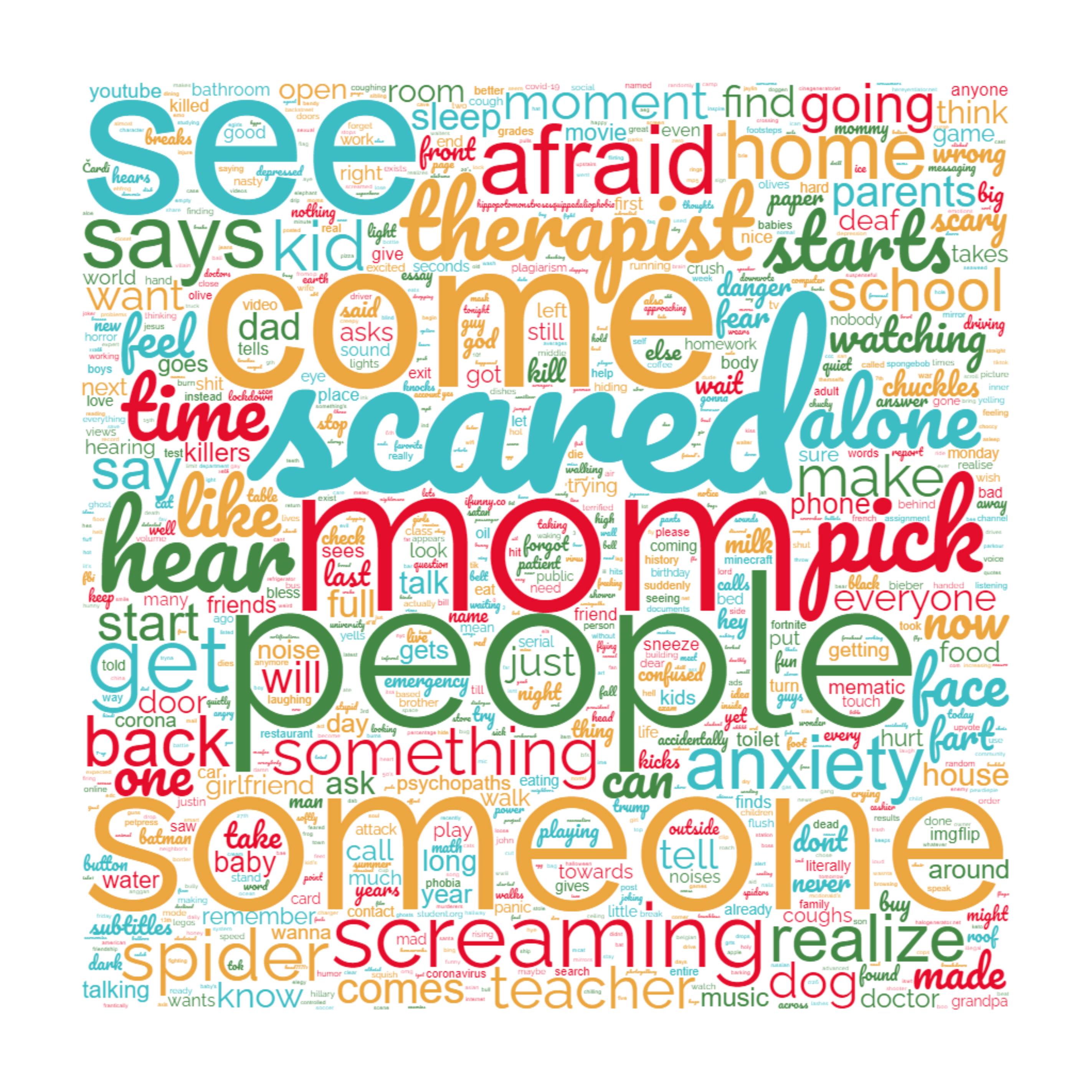}}
    \subfigure[Anger]{\includegraphics[width=0.16\textwidth]{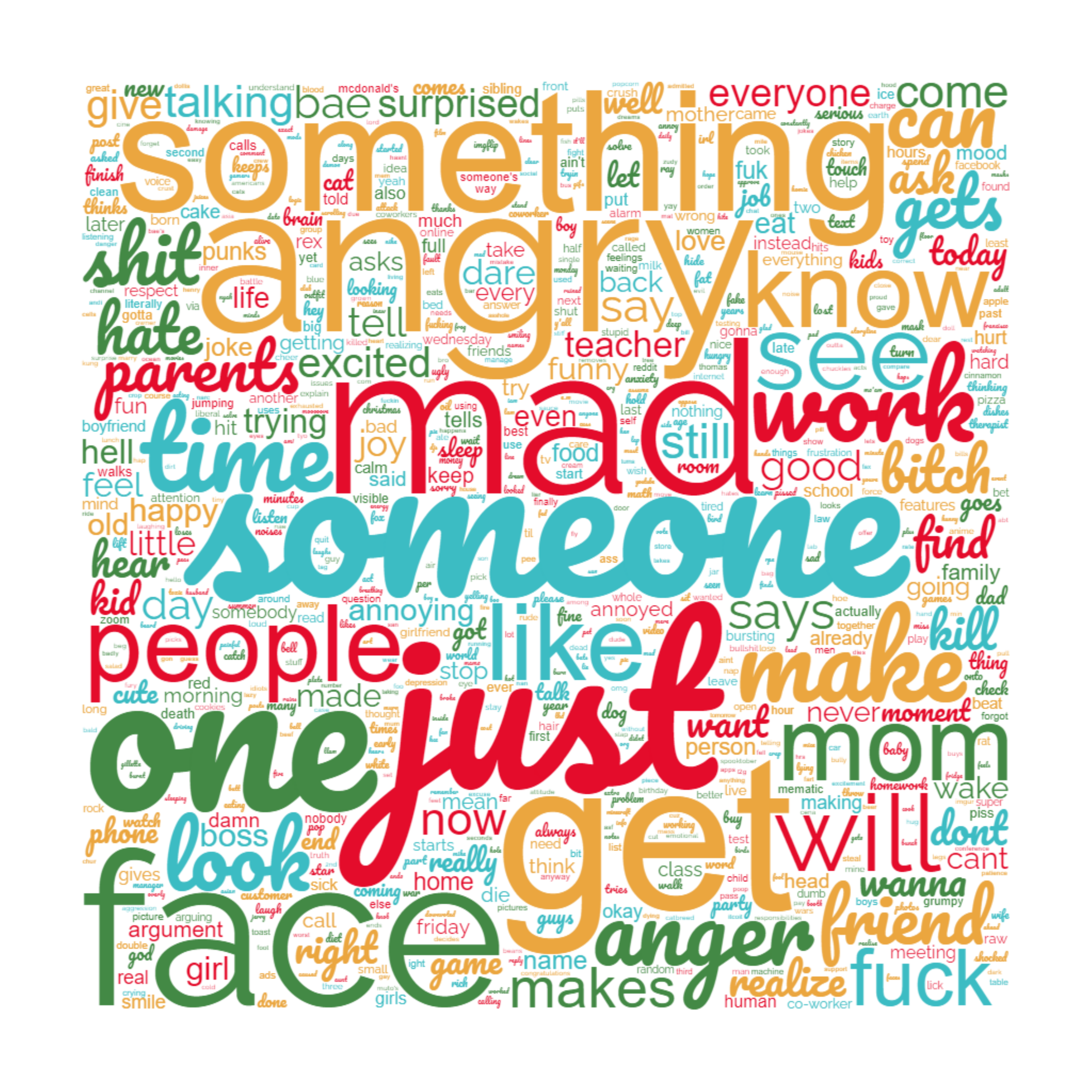}} 
    \subfigure[Joy]{\includegraphics[width=0.16\textwidth]{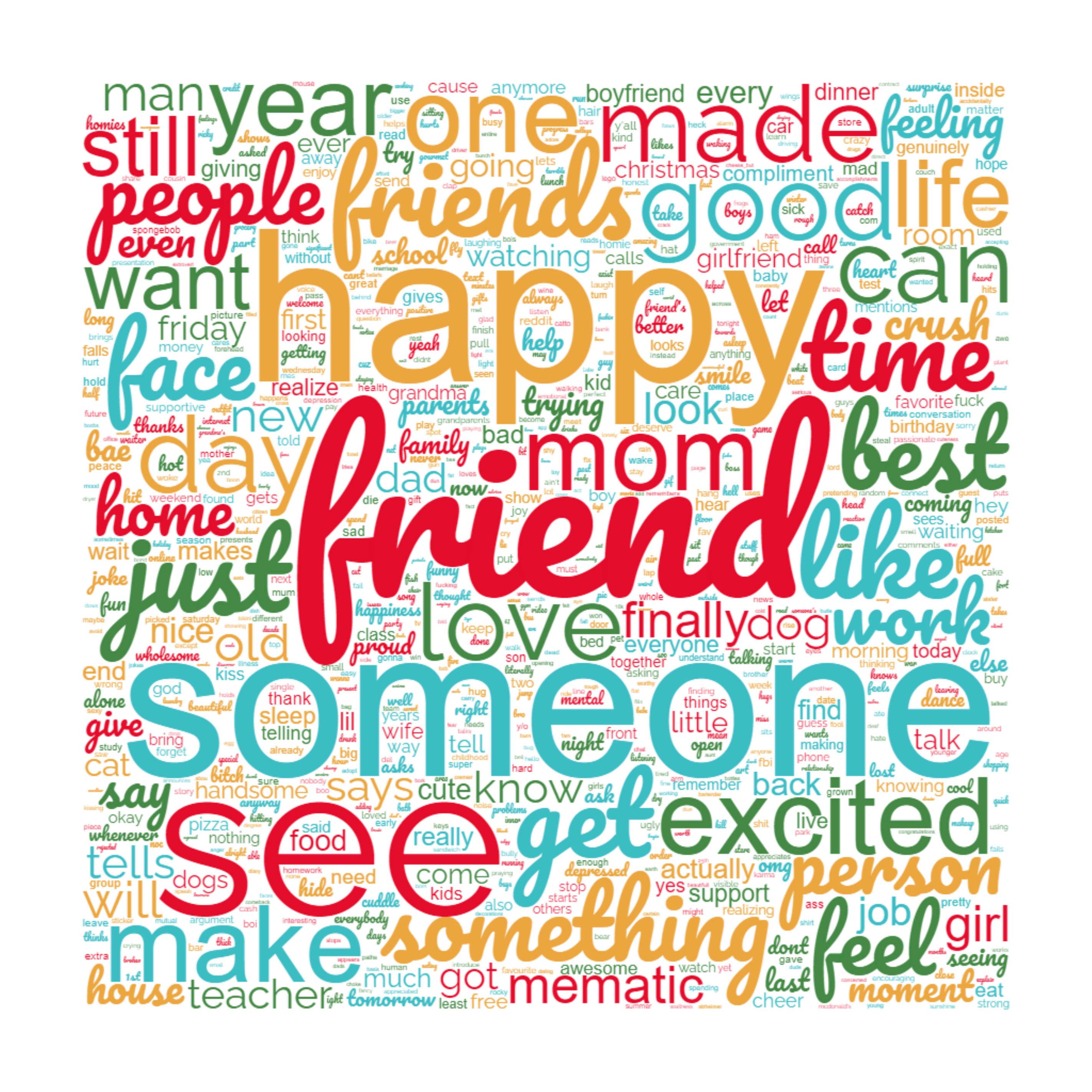}}
    \subfigure[Sadness]{\includegraphics[width=0.16\textwidth]{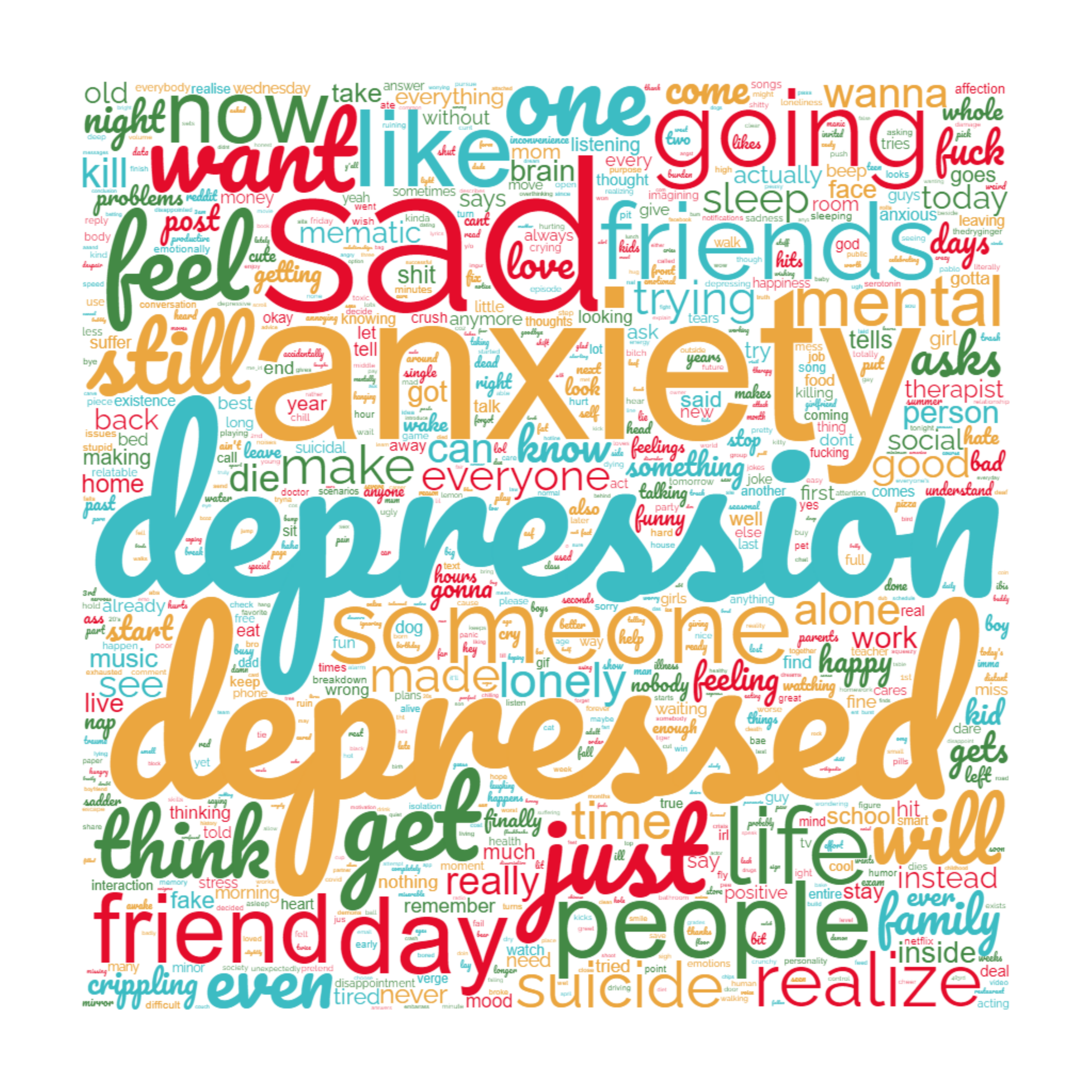}}
    \subfigure[Surprise]{\includegraphics[width=0.16\textwidth]{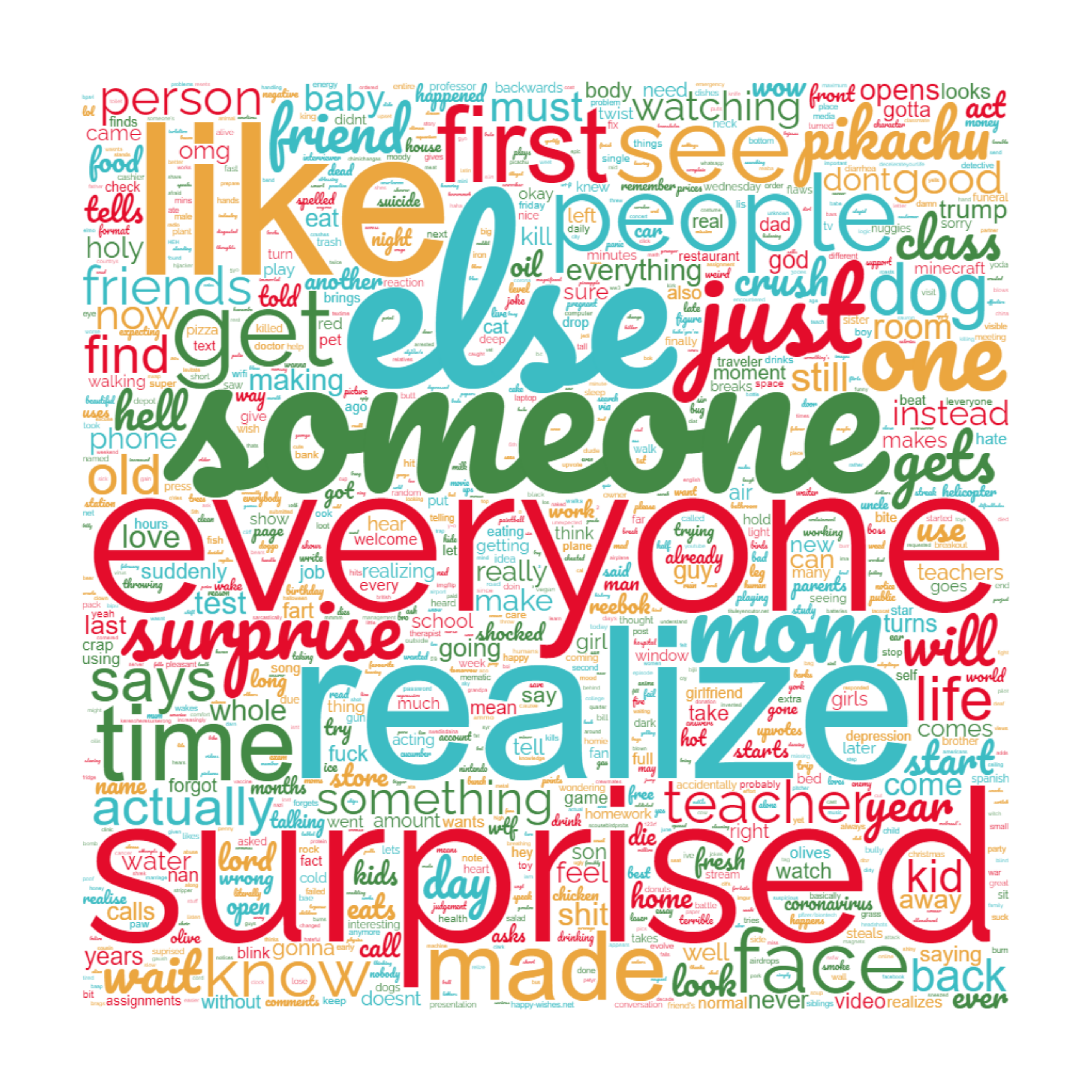}}    
    \subfigure[Disgust]{\includegraphics[width=0.16\textwidth]{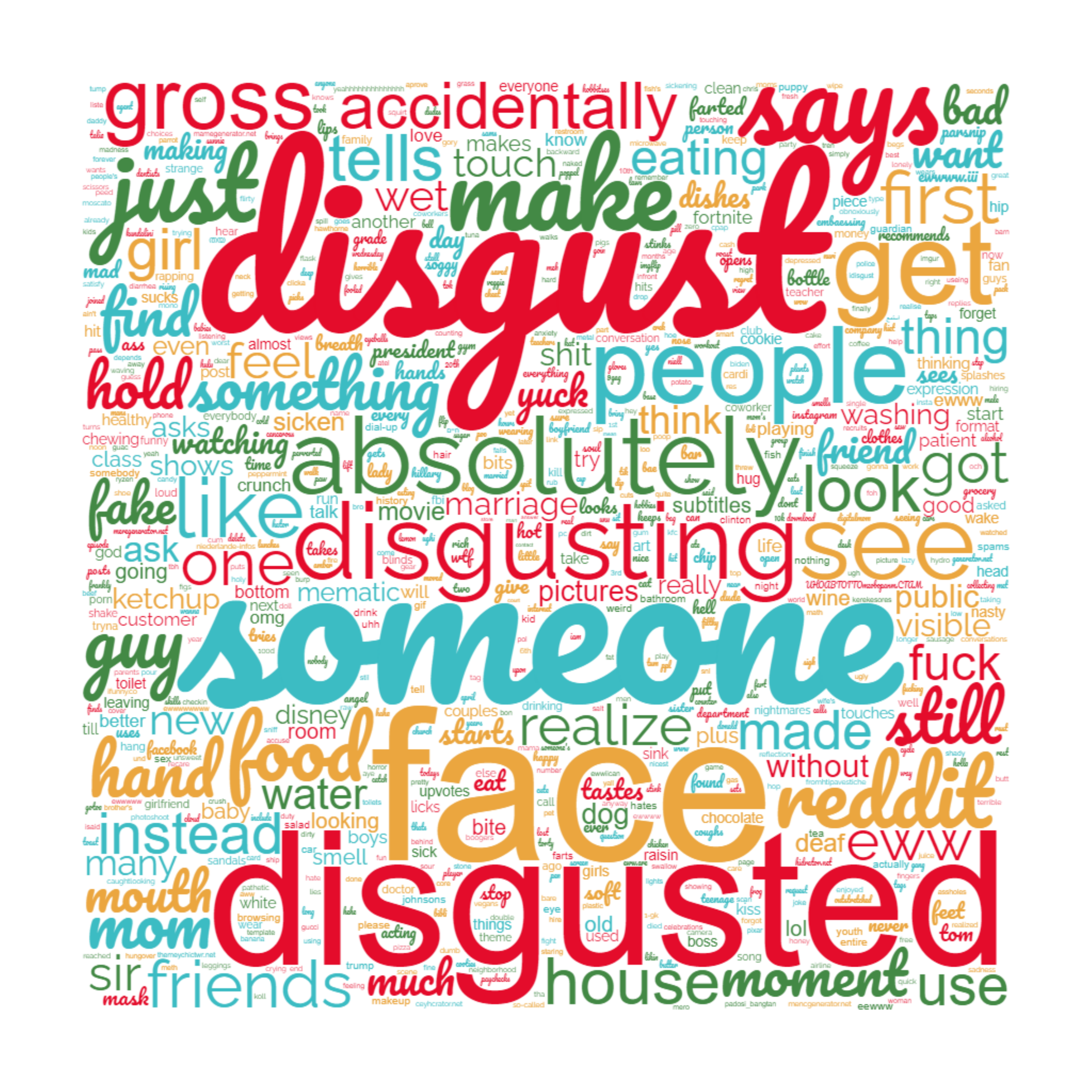}} 
    \caption{Word clouds depicting the category-wise lexicon comprising the embedded texts for memes in the \dataset\ dataset.}
    \label{fig:wordclouds}
    \vspace{-5mm}
\end{figure*}

\subsection{Lexical Analysis of \dataset}
\label{sec:lex_sum}
The lexical summary exhibited by the textual cues present in the memes, in the form of overlaid text (c.f. Fig. \ref{fig:wordclouds}), suggests interesting characteristics. All categories except \textit{surprise}, prominently exhibit affect-enriched lexicon, including nouns, adjectives and verbs, exemplified as: \textit{disgust} \texttt{(gross, yuck, eww)}, \textit{anger} \texttt{(punks, hate, mad)}, \textit{fear} \texttt{(afraid, screaming, therapist)}, \textit{sadness} \texttt{(depressed, anxiety, lonely)} and \textit{joy} \texttt{(friend, happy, love)}. Whereas, as elucidated from Fig. \ref{fig:wordclouds} (e), \textit{surprise}-based emotion characterization relies significantly upon contextual cues, instead of lexical ones.

Additionally, Table \ref{tab:top-words} (c.f. Appendix \ref{app:subsec:themes}) shows top 10 frequent words in the meme text for each category after masking the `category-keywords'. The corresponding TF-IDF scores are given in the parenthesis. Some category-specific relevant words can be observed at the top of the category columns, except for \textit{anger}. We do not see any particular word that is usually used in the contexts conveying \textit{anger}. This suggests the complex constructs people typically use while creating memes when conveying \textit{anger}. Essentially, the anger might not be conveyed using explicit keywords but by using implicit and complex references instead. This indicates the complexity the affective content from social media can pose to multimodal systems. Fig. \ref{fig:text_length_dist} shows the length distribution of meme text for each class; we observe no significant differences between the classes.

\begin{figure}[t!]
\centering
\includegraphics[width=\columnwidth]{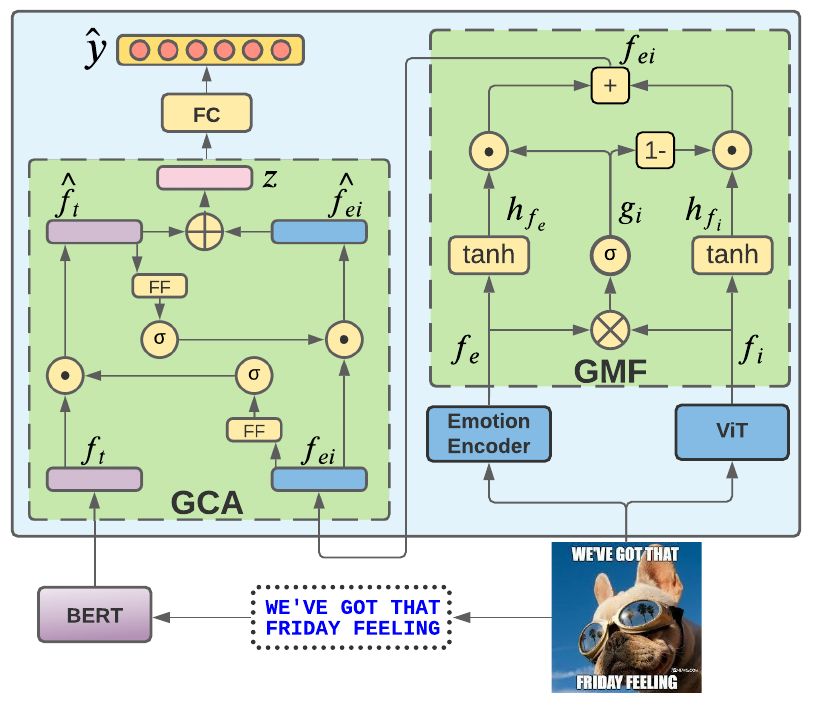}
\caption{\proposed's model architecture. GCA: Gated Cross Attention module, GMF: Gated Multimodal Fusion module, $\otimes$: Low-rank Bilinear Pooling, $\oplus$: Concatenation.}
\label{fig:proposed_arch}
\end{figure}

\section{Proposed Approach}
As shown in Fig. \ref{fig:proposed_arch}, our proposed approach utilizes the meme image and embedded text, extracted using Google GCV OCR\footnote{\href{https://cloud.google.com/vision/docs/ocr}{Google Cloud Vision OCR API} \ExternalLink} (GOCR) as primary inputs. For encoding image features, we use Vision Transformer (ViT) \cite{dosovitskiy2021image} and for meme text, we use BERT \cite{devlin2019BERT}.

We model emotion-aware image representations by explicitly incorporating visually depicted emotions. We extract emotion features from the images using a ViT-based encoder network, pre-trained using the \textit{extended} AffectNet dataset. 
We then pass the emotion and image features through a \textit{gated} multimodal unit to obtain an emotion-aware image representation, as shown in Fig. \ref{fig:proposed_arch}. Further, we feed the emotion-aware image and textual representation through a gated cross-modal attention module that facilitates selective cross-modal attuning and blocking. Finally, we concatenate the two resultant updated representations and pass the joint representation of the meme through a feed-forward network toward the classification task. 
We describe each of these modules in detail below.

\subsection{Unimodal Feature Extraction}

We use pre-trained unimodal encoders to obtain representations of the image ($i$) and text ($t$) for a given meme ($M$), as described below. 

\noindent \textbf{-- Image Encoder:} We use ViT \cite{dosovitskiy2021image} initialised with ImageNet weights as the image encoder and obtain an $m\times768$ dimensional output corresponding to $m$ image patches, i.e., $\mathbf{f_i} = \text{ViT}(i) \in\mathbb{R}^{m\times768}$.

\noindent \textbf{-- Text Encoder:} We use pre-trained BERT \cite{devlin2019BERT} as the text encoder. Specifically, we take the token-level representations corresponding to $n$ tokens from the last hidden layer: $\mathbf{f_t} = \text{BERT}(t) \in\mathbb{R}^{n\times768}$. 

\subsection{Emotion Feature Extraction}
\label{subsec:emopretrain}
There are well-established variants of approaches, specially tailored towards capturing semantic objects and salient features, like RCNNs \cite{rfcn,maskrcnn}, YOLO \cite{yolov7}, and CenterNet \cite{obaspoints,cornernet}, to name a few; yet, there is a dearth of solutions addressing complex emotion features from visuals, especially memes. Towards explicitly incorporating visually-depicted emotion features as part of our proposed methodology, we build an emotion feature extraction model by fine-tuning a ViT-based image patch encoder for the emotion expression classification task. We then freeze its weights toward extracting relevant emotion-enriched cues from a given meme. To this end, we leverage the AffectNet dataset, a large-scale dataset of typical human facial expressions; but we also extend it by adding non-human subjects.

After pretraining a ViT base model using over $50K$ emotion-enriched images from the \textit{extended} AffectNet dataset for emotion classification, we freeze its weights for extracting emotion-enriched image features. These features are incorporated as part of \proposed.\ This is expressed as $\mathbf{f_{e}} = \text{ViT}(i) \in\mathbb{R}^{m\times768}$.

\subsection{Gated Multimodal Fusion (GMF)}
As part of effectively incorporating visual cues from memes for meme emotion recognition, it is crucial to optimally infuse emotion-enriched input signals while emphasizing other relevant visual cues. This becomes critical while fusing features from similar source modalities.  

To induce selective processing of input features, we adapt a \textit{gated multimodal unit} \cite{arevalo2017gated} by using low-rank bilinear pooling (LRBP) \cite{kim2017hadamard} while computing a sigmoid-based gating weight, instead of a simple concatenation based approach. The motivation for this change is the requirement to fuse the representations coming from the same input source (i.e., image), towards which a Hadamard product-based interaction is empirically observed to be preferable over a concatenation-based approach. This module performs a fusion of the emotion features ($\mathbf{f_{e}}$) that are extracted using the emotion encoder, and the meme image features ($\mathbf{f_{i}}$) obtained using the image encoder to finally obtain emotion-aware image features ($\mathbf{f_{ei}}$). We do this as we have two different types of image encodings, $\mathbf{f_{e}}$ and $\mathbf{f_{i}}$. Such a fusion trades off on how much novel information is required from each encoding using a sigmoid-based gated fusion mechanism. 
\begin{small}
\begin{align}
    \mathbf{h_{f_i}} & = \tanh(\mathbf{f_{i}\mathbf{W_{i}}} + \mathbf{b_{i}}); \quad & \mathbf{h_{f_e}} & = \tanh(\mathbf{f_{e}}\mathbf{W_{e}} + \mathbf{b_{e}})  \\
    \mathbf{g_{i}} & = \sigma([\mathbf{h_{f_i}} \otimes \mathbf{h_{f_e}}]\mathbf{W_{g}}); \quad & \mathbf{f_{ei}} & = \mathbf{g_{i}h_{f_e}}+(1-\mathbf{g_{i}})\mathbf{h_{f_i}}
\end{align}
\end{small}
\noindent where $\mathbf{W_{i}}$, $\mathbf{W_{e}}$, $\mathbf{W_{g}}$ $\in\mathbb{R}^{768\times768}$ are the weights for transforming image features ($\mathbf{f_{i}}\in\mathbb{R}^{m\times768}$), emotion features ($\mathbf{f_{e}}\in\mathbb{R}^{m\times768}$), and low-rank bilinear pooling based fusion ($\mathbf{g_{i}}\in\mathbb{R}^{m\times768}$) of their latent features ($[\mathbf{h_{f_i}}\otimes \mathbf{h_{f_e}}]\in\mathbb{R}^{m\times768}$), respectively. The bias terms, $\mathbf{b_{i}}$ and $\mathbf{b_{e}}$ $\in\mathbb{R}^{768}$,   correspond to the representation learning for image and its emotion-enriched signals, respectively. $\sigma$ denotes the \textit{sigmoid} activation function. Also, numpy-like broadcasting is inherently applied wherever applicable via PyTorch API\footnote{\href{https://pytorch.org/docs/stable/notes/broadcasting.html}{Broadcasting Semantics | PyTorch \ExternalLink}}.

\begin{table*}[!htb]
\caption{Comparison of \proposed\ and baselines on the \dataset\ dataset.  The last row shows the improvement of  \proposed\ ($^{*}$) over the \textit{early-fusion} based model, designated as the best baseline ($\dagger$). Class-wise accuracy for the six \textit{Ekman} emotions in \dataset\ is also reported.}
\resizebox{\textwidth}{!}{
\centering
\begin{tabular}{clcccccccccc}
\toprule
Modality &
  \multicolumn{1}{c}{Model} &
  \multicolumn{1}{c}{Acc.} &
  \multicolumn{1}{c}{Prec.} &
  \multicolumn{1}{c}{Rec.} &
  \multicolumn{1}{c}{F1} &
  \multicolumn{1}{c}{FER} &
  \multicolumn{1}{c}{AGR} &
  \multicolumn{1}{c}{JOY} &
  \multicolumn{1}{c}{SDN} &
  \multicolumn{1}{c}{SPR} &
  \multicolumn{1}{c}{DGT} \\ \cmidrule(lr){1-2}\cmidrule(lr){3-6}\cmidrule(lr){7-12}
\multirow{2}{*}{UM}                        & BERT                      & 0.633  & 0.6537 & 0.6337 & 0.6387 & 0.4573 & 0.5587 & \textbf{0.7969} & 0.7484 & 0.4559 & 0.6869 \\ 
                      & ViT                       & 0.6713 & 0.6913 & 0.6713 & 0.6738 & \textbf{0.9065} & 0.5507 & 0.6884 & 0.6243 & 0.8766 & 0.7101 \\ \midrule
\multirow{6}{*}{MM} & Early-fusion$^{\dagger}$              & 0.7836 & 0.8121 & 0.7836 & 0.7749 & 0.8991 & 0.7594 & 0.7405 & 0.657  & 0.8335 & 0.833  \\
                            & MMBT                      & 0.6337 & 0.6537 & 0.633  & 0.6352 & 0.581  & \textbf{0.7848} & 0.7818 & 0.6534 & 0.5252 & 0.7973 \\ 
                            & CLIP                      & 0.6378 & 0.8027 & 0.6378 & 0.6816 & 0.5351 & 0.5011 & 0.7797 & \textbf{0.764}  & 0.5226 & 0.72   \\ 
                            & VisualBERT               & 0.6725 & 0.7961 & 0.6725 & 0.7002 & 0.7075 & 0.5838 & 0.6969 & 0.7294 & 0.7787 & 0.5513 \\ \cmidrule{2-12} 
                            & \proposed$^{\star}$                  & \textbf{0.8239} & \textbf{0.8314} & \textbf{0.8239} & \textbf{0.8243} & 0.8777 & 0.7835 & 0.7924 & 0.7625 & \textbf{0.8935} & \textbf{0.8392} \\ \midrule
\multicolumn{2}{c}{$\Delta_{\text{($\star$-$\dagger$)}\times 100}(\%)$} & \textcolor{blue}{$\uparrow4.03\%$} & \textcolor{blue}{$\uparrow1.93\%$}& \textcolor{blue}{$\uparrow4.03\%$}& \textcolor{blue}{$\uparrow4.94\%$}& \textcolor{red}{$\downarrow2.14\%$}& \textcolor{blue}{$\uparrow2.41\%$}& \textcolor{blue}{$\uparrow5.19\%$} & \textcolor{blue}{$\uparrow10.55\%$}& \textcolor{blue}{$\uparrow6.00\%$} & \textcolor{blue}{$\uparrow0.62\%$}\\\bottomrule 
\end{tabular}}
\label{tab:task-0-combined}
\vspace{-5mm}
\end{table*}

\subsection{Gated Cross Attention (GCA)}
Prominent conventional approaches that leverage co-attentional transformers-based layers have been observed to perform well in scenarios involving visual-linguistic grounding \cite{lu2019vilbert,li2019visualbert}. However, they exhibit sub-optimal results while modeling memes \cite{pramanick-etal-2021-momenta-multimodal}. This could be likely due to the cross-modal noise being captured and attended to while learning dissociated cues from modality-specific meme components.  
Towards regulating the inherent effect of cross-modal noise, we modify the cross-attention mechanism \cite{cross_attn}, by incorporating the adaptive co-attention strategy \cite{adaptive_co_attn}. Instead of incorporating self-attention layers for cross-modal attention, we perform gating over one modality (visual) first, followed by weighting the other modality (textual). We then perform gated attention for the first modality (visual) using the weighted textual representation to obtain its feedback-based representation. We call this \textit{Gated Cross Attention} mechanism. It facilitates the extraction of useful features from emotion-aware image ($\mathbf{f_{ei}}\in\mathbb{R}^{m\times768}$) and textual ($\mathbf{f_{t}}\in\mathbb{R}^{n\times768}$) features. Thus, we obtain new feature representations, $\mathbf{\hat{f_{ei}}}\in\mathbb{R}^{m\times768}$ and $\mathbf{\hat{f_t}}\in\mathbb{R}^{m\times768}$, as follows:

\begin{small}
\begin{equation} \label{eq1}
\mathbf{h_{f_{ei}}}  = \sigma(\mathbf{f_{ei}}\mathbf{W_{ei}} + \mathbf{b_{ei}}); \quad \mathbf{\alpha_{ei}}  = \text{softmax}(\mathbf{h_{f_{ei}}}\mathbf{W_{\alpha_{ei}}} + \mathbf{b_{\alpha_{ei}}})
\end{equation}
\begin{equation} \label{eq2}
\begin{split}
\mathbf{\hat{f_t}}  = \mathbf{\alpha_{ei}} \times \mathbf{f_{t}} \in\mathbb{R}^{m\times768}
\end{split}
\end{equation}
\begin{equation} \label{eq3}
\mathbf{h_{\hat{f_t}}}  = \sigma(\mathbf{\hat{f_{t}}}\mathbf{W_{t}} + \mathbf{b_{t}}); \quad  \mathbf{\alpha_{t}}  = \text{softmax}(\mathbf{h_{\hat{f_t}}}\mathbf{W_{\alpha_t}} + \mathbf{b_{\alpha_t}})
\end{equation}
\begin{equation} \label{eq4}
\begin{split}
\mathbf{\hat{f_{ei}}} & = \mathbf{\alpha_{t}} \times \mathbf{f_{ei}} \in\mathbb{R}^{m\times768}
\end{split}
\end{equation}
\end{small}

where $\mathbf{W_{ei}}$,  $\mathbf{W_{t}}$  $\in\mathbb{R}^{768\times768}$, $\mathbf{W_{\alpha_{ei}}}$, and $\mathbf{W_{\alpha_t}}$  $\in\mathbb{R}^{768\times1}$ are the weights for transforming emotion-aware image feature ($\mathbf{f_{ei}}\in\mathbb{R}^{m\times768}$) repeated $n$ times to account for $n$ textual tokens), updated text feature ($\mathbf{\hat{f_{t}}}\in\mathbb{R}^{m\times768}$) repeated $m$ times to account for $m$ image patches), intermediate representation of transformed emotion-aware image feature ($\mathbf{h_{f_{ei}}}\in\mathbb{R}^{m\times n\times768}$), and intermediate representation of transformed text feature ($\mathbf{h_{\hat{f_t}}}\in\mathbb{R}^{m\times m\times768}$), respectively. The bias terms, $\mathbf{b_{ei}}$, $\mathbf{b_{t}}$ $\in\mathbb{R}^{768}$, and $\mathbf{b_{\alpha_{ei}}}$, $\mathbf{b_{\alpha_t}}$ $\in\mathbb{R}^{1}$,  correspond to the representation learning for $\mathbf{h_{f_{ei}}}, \mathbf{h_{\hat{f_t}}}, \mathbf{\alpha_{ei}}$, and $\mathbf{\alpha_t}$, respectively. 

\subsection{Prediction and Training Objective}

Finally, we apply sum-pooling across the \textit{first} dimension of the corresponding weight-aggregated features: $\mathbf{\hat{f_{ei}}}\in\mathbb{R}^{m\times768}$ and $\mathbf{\hat{f_t}}\in\mathbb{R}^{m\times768}$, followed by concatenating the sum-pooled features ($\mathbf{\hat{f_{ei}}}\in\mathbb{R}^{768}$ and $\mathbf{\hat{f_t}}\in\mathbb{R}^{768}$), to produce a joint meme representation ($\mathbf{f_{z_1}}\in\mathbb{R}^{1536}$). This is given as input to a feed-forward network for the final classification. 
\begin{small}
\begin{align}
    \mathbf{f_{z_1}} &= [sum(\mathbf{\hat{f_{ei}}), sum(\mathbf{\hat{f_t}})] \in\mathbb{R}^{1536}} \\ 
    \mathbf{h_{f_{z_1}}} &  = \text{relu}(\mathbf{f_{z_1}}\mathbf{W_{z_1}} + \mathbf{b_{z_1}}) \in\mathbb{R}^{768} \\
    \mathbf{h_{f_{z_2}}} & = \text{softmax}(\mathbf{h_{f_{z_1}}}\mathbf{W_{z_2}} + \mathbf{b_{z_2}}) \in\mathbb{R}^{6} \\
    \mathbf{\hat{y}} &= argmax_{\mathbf{y}\in \mathbf{\mathcal{Y}}}f(\mathbf{y};\mathbf{\theta}, \mathbf{h_{f_{z_2}}})
\end{align}
\end{small}

where $\mathbf{W_{z_1}}\in\mathbb{R}^{1536\times768}$, $\mathbf{W_{z_2}}\in\mathbb{R}^{768\times6}$, $\theta$ represents model parameters, $\mathbf{\hat{y}}$ is the predicted class index, and $\mathbf{\mathcal{Y}}$ is the label-id set, with $|\mathbf{\mathcal{Y}}|=6$. The bias terms, $\mathbf{b_{z_1}}\in\mathbb{R}^{768}$ and $\mathbf{b_{z_2}}\in\mathbb{R}^{6}$, are corresponding to the last two layers.
We use the cross-entropy loss for optimization. Moreover, we employ the online label smoothing \cite{ols} for regularization.

\section{Baseline Models}

\noindent \textbf{Unimodal Baselines:}
We restrict modality-specific encoders to the following choices, as the primary objective of this work is to investigate an optimal multimodal fusion strategy. 
\begin{itemize}[leftmargin=*]
    \item \textbf{BERT \cite{devlin2019BERT}:} We use the BERT-base-uncased model as our text-only baseline.
    \item \textbf{ViT \cite{dosovitskiy2021image}:} We use ViT with ImageNet weights as our image-only baseline.
\end{itemize}

\noindent \textbf{Multimodal Baselines:}
For multimodal systems, we explore the following competent approaches as comparative baselines. These systems endorse various multimodal interaction schemes, facilitating a robust assessment. 

\begin{itemize}[leftmargin=*]
    \item \textbf{Early-fusion:} In this model, the features from ViT and BERT are concatenated and passed through a feed-forward network for classification.
    \item \textbf{MMBT \cite{kiela2020mmbt}:} It is a supervised bi-modal transformer that projects image features from unimodally pre-trained image encoders to text tokens. 
    \item \textbf{CLIP \cite{radford2021clip}:} CLIP is a contrastive learning-based approach that is designed to learn visual information through natural language supervision. 
    \item \textbf{VisualBERT \cite{li2019visualbert}:} VisualBERT is a transformer-based model for visuo-lingual modelling. It has been trained on the MS COCO dataset employing masked language modeling and the sentence-image prediction objective functions.
\end{itemize}

\section{Experiments}
Firstly, we compare \proposed\ with both unimodal (image/text) and multimodal models. The comparisons are first made for meme emotion detection task using the \dataset\ dataset, followed by evaluation on three \memotion\ tasks \cite{sharma2020semeval2020} (c.f. Section \ref{sec:memotion}). We further perform the ablation of our model. This is followed by examining the interpretability of \proposed\ using GradCAM \cite{gradcam}. We then demonstrate the generalizability of \proposed's\ performance on the  \harmeme\ \cite{harmeme} and \dank\ datasets \cite{dank_evalita}. Finally, we analyze the errors observed during performance evaluation. Our experimental setup's empirical examinations involve fine-tuning for the respective tasks and datasets. Since we primarily explore a multi-class classification setup, we are more interested in evaluating the system performances that factors-in the class-wise contributions equally. Therefore, we use macro-averaged formulations of accuracy, precision, recall, and F1-score as evaluation metrics. We also report class-wise F1

\subsection{Implementation Details and Hyperparameter Values}
\label{sec:hyper}
\begin{table}[t!]
\caption{Hyperparameters of different models.}
\centering
\resizebox{\columnwidth}{!}
{
\begin{tabular}{ccccccc}
\toprule
Model & BS & Ep & LR & Image Encoder & Text Encoder & \# Param \\ \midrule
TextBERT           & 64 & 20 & 0.0001   & -           & bert-base-uncased & 110M \\ 
ViT & 32 & 20 & 0.0001   & vit-base    & -                 & 86M  \\ 
Early-fusion       & 32 & 20 & 0.0001   & vit-base    & bert-base-uncased & 196M \\ 
CLIP               & 43 & 20 & 0.0001   & vit-base    & bert-base-uncased & 151M \\ 
MMBT               & 32 & 30 & 0.00001  & resnet152   & bert-base-uncased & 169M \\ 
V-BERT        & 32 & 30 & 0.000001 & Faster RCNN & bert-base-uncased & 247M \\ 
\proposed           & 32 & 30 & 0.0001   & vit-base    & bert-base-uncased & 282M \\ \bottomrule
\end{tabular}}
\label{tab:hyperparams-appendix}
\vspace{-5mm}
\end{table}
We train all the models using Pytorch 1.10 on an Nvidia Tesla V100 GPU with 32 GB dedicated memory, CUDA-11.2 and cuDNN-8.1.1 installed. For the primary emotion classification, \memotion,\ \harmeme\ tasks, we use BERT as the text encoder and ViT as the image encoder. Specifically, we use the bert-base-uncased checkpoint for BERT and google/vit-base-patch16-224 checkpoint for ViT. However, for the \dank\ task, we switch BERT with UmBERTo, which is a BERT-based model but pre-trained using Italian corpus. The linear layers in GCA and GMF modules are initialized using Xavier initialization, and the bias is set to zero. For the meme emotion classification task, we train all the models using the \textit{online label smoothing} loss and Adam optimizer. For the \memotion,\ \harmeme\ and \dank\ tasks, we use cross-entropy loss and Adam optimizer. We also present these details in Table \ref{tab:hyperparams-appendix}.

\subsection{Evaluation on the \dataset\ dataset}
Among unimodal models, the image-only model is observed to perform better (c.f. Table \ref{tab:task-0-combined}) than the text-only model by $4\%$ F1 score. Also, multimodal baselines are observed to perform either at par or better than unimodal models. One of the top-performing baselines, as shown in Table \ref{tab:task-0-combined}, is the early-fusion model with BERT and ViT as its text and image encoders, respectively. This could be due to an optimal modeling requirement posed by the meme emotion detection task, which does not seem to favor complex co-attentive visual-linguistic grounding employed by models like MMBT, CLIP, and VisualBERT.

The early-fusion model ($0.7749$) yields a $7\%$ absolute improvement in F1-score over the sophisticated VisualBERT ($0.7002$). Overall, both perform better than the multimodal baselines like MMBT ($0.6352$) and CLIP ($0.6816$). 
In comparison, \proposed\ registers $4.94\%$ F1 improvement over the early-fusion model. This improvement could be mainly attributed to the GMF-based explicit emotion modeling and GCA-based inter-modal fusion, facilitating preferential treatment for both input modalities conditioned upon emotion-enriched visual cues. Overall,  \proposed\ yields an improvement of 1.93\%-4.94\% across all four metrics. 

\proposed\ significantly increases accuracy for four classes -- \textit{anger} ($\uparrow$2.41\%), \textit{joy} ($\uparrow$5.19\%), \textit{sadness} ($\uparrow$10.55\%) and \textit{surprise} ($\uparrow$6.00\%). In contrast, the accuracy for \textit{disgust} improves slightly ($0.62\%$), but not as much as for the classes above. This subtle enhancement observed could be likely due to the expressiveness of the emotion \textit{disgust} via either text, image, or even both (See Fig. \ref{fig:six_meme_eg} (a)). Whereas the lower representation of the \textit{disgust} class in the dataset explains the improvement that is minor compared to that of the categories mentioned above. Moreover, the performance for \textit{fear} drops by $2\%$, wherein the discriminatory cues are predominantly image-based, the implication of which is also corroborated by the highest category-specific performance ($\approx0.91$ F1-score), by ViT-based model. Besides posing challenges like object occlusion, complex pose, image quality, etc., the visual modeling is impacted by the category's under-representation in both \dataset\ and extended AffectNet datasets, leading to \proposed's drop in accuracy for \textit{fear} as against the enhancement observed for other categories.

\begin{table*}[htb!]
\caption{Task-wise and class-wise performance (Macro-F1) of different approaches on the \memotion\ tasks. The last row shows the improvement of \proposed\ ($^{*}$) over the {\em previous best} results ($\dagger$) reported.}
\centering
\resizebox{\textwidth}{!}{
\begin{tabular}{clccccccccccc}
\toprule
\multicolumn{1}{c}{\multirow{2}{*}{Modality}} &
\multicolumn{1}{c}{\multirow{2}{*}{Model}} &
  \multicolumn{1}{c}{\sent} &
  \multicolumn{5}{c}{\emot} &
  \multicolumn{5}{c}{\emotq} \\ \cmidrule(lr){3-3}\cmidrule(lr){4-8}\cmidrule(lr){9-13} 
\multicolumn{1}{c}{} & &
  Sentiment &
  \multicolumn{1}{c}{Humour} &
  \multicolumn{1}{c}{Sarcasm} &
  \multicolumn{1}{c}{Offensive} &
  \multicolumn{1}{c}{Motivation} &
  \multicolumn{1}{c}{\textbf{Average}} &
  \multicolumn{1}{c}{Humour} &
  \multicolumn{1}{c}{Sarcasm} &
  \multicolumn{1}{c}{Offense} &
  \multicolumn{1}{c}{Motivation} &
  \textbf{Average} \\ \midrule
  {\multirow{2}{*}{UM}} & BERT &
  0.3123 &
  \multicolumn{1}{c}{0.5235} &
  \multicolumn{1}{c}{0.4747} &
  \multicolumn{1}{c}{0.5012} &
  \multicolumn{1}{c}{0.5089} &
  0.5021 &
  \multicolumn{1}{c}{0.2487} &
  \multicolumn{1}{c}{0.2309} &
  \multicolumn{1}{c}{0.2333} &
  \multicolumn{1}{c}{0.4641} &
  0.2942 \\ 
& ViT &
  0.3158 &
  \multicolumn{1}{c}{0.5114} &
  \multicolumn{1}{c}{0.4851} &
  \multicolumn{1}{c}{0.5119} &
  \multicolumn{1}{c}{0.519} &
  0.5069 &
  \multicolumn{1}{c}{0.2434} &
  \multicolumn{1}{c}{0.249} &
  \multicolumn{1}{c}{0.2388} &
  \multicolumn{1}{c}{0.4972} &
  0.307 \\ \midrule
\multirow{6}{*}{MM} & Early-fusion &
  0.3295 &
  \multicolumn{1}{c}{0.5122} &
  \multicolumn{1}{c}{0.5032} &
  \multicolumn{1}{c}{0.5059} &
  \multicolumn{1}{c}{0.4591} &
  0.4951 &
  \multicolumn{1}{c}{0.2368} &
  \multicolumn{1}{c}{0.2426} &
  \multicolumn{1}{c}{0.235} &
  \multicolumn{1}{c}{0.4481} &
  0.2906 \\
& MMBT &
  0.3457 &
  \multicolumn{1}{c}{\textbf{0.5393}} &
  \multicolumn{1}{c}{0.5015} &
  \multicolumn{1}{c}{0.4970} &
  \multicolumn{1}{c}{0.4989} &
  0.5092 &
  \multicolumn{1}{c}{0.2474} &
  \multicolumn{1}{c}{0.2364} &
  \multicolumn{1}{c}{0.2475} &
  \multicolumn{1}{c}{0.4838} &
  0.3038 \\ 
& CLIP &
  0.3261 &
  \multicolumn{1}{c}{0.4798} &
  \multicolumn{1}{c}{0.5133} &
  \multicolumn{1}{c}{0.5115} &
  \multicolumn{1}{c}{0.4967} &
  0.5003 &
  \multicolumn{1}{c}{0.2652} &
  \multicolumn{1}{c}{\textbf{0.2549}} &
  \multicolumn{1}{c}{0.2448} &
  \multicolumn{1}{c}{0.4754} &
  0.3101 \\
 & Previous Best$^{\dagger}$ &
  \multicolumn{1}{c}{\textbf{0.3547}} &
  \multicolumn{1}{c}{0.51587} &
  \multicolumn{1}{c}{0.5159} &
  \multicolumn{1}{c}{\textbf{0.5225}} &
  \multicolumn{1}{c}{0.51909} &
  \multicolumn{1}{c}{0.5183} &
  \multicolumn{1}{c}{\textbf{0.27069}} &
  \multicolumn{1}{c}{0.25028} &
  \multicolumn{1}{c}{0.25761} &
  \multicolumn{1}{c}{0.51126} &
  0.3225 \\ \cmidrule{2-13}
& \proposed$^{\star}$ &
  \multicolumn{1}{c}{0.3486} &
  \multicolumn{1}{c}{0.5268} &
  \multicolumn{1}{c}{\textbf{0.5272}} &
  \multicolumn{1}{c}{0.5175} &
  \multicolumn{1}{c}{\textbf{0.5375}} &
  \textbf{0.5272} &
  \multicolumn{1}{c}{0.2646} &
  \multicolumn{1}{c}{0.2501} &
  \multicolumn{1}{c}{\textbf{0.2598}} &
  \multicolumn{1}{c}{\textbf{0.5275}} &
  \textbf{0.3253} \\ \midrule
\multicolumn{2}{c}{$\Delta_{\text{($\star$-$\dagger$)}\times 100}(\%)$} &
  \multicolumn{1}{c}{\textcolor{red}{$\downarrow0.61\%$}} &
  \multicolumn{1}{c}{\textcolor{blue}{$\uparrow1.09\%$}} &
  \multicolumn{1}{c}{\textcolor{blue}{$\uparrow1.13\%$}} &
  \multicolumn{1}{c}{\textcolor{red}{$\downarrow0.5\%$}} &
  \multicolumn{1}{c}{\textcolor{blue}{$\uparrow1.84\%$}} &
  \multicolumn{1}{c}{\textcolor{blue}{$\uparrow0.89\%$}} &
  \multicolumn{1}{c}{\textcolor{red}{$\downarrow0.6\%$}} &
  \multicolumn{1}{c}{\textcolor{red}{$\downarrow0.01\%$}} &
  \multicolumn{1}{c}{\textcolor{blue}{$\uparrow0.22\%$}} &
  \multicolumn{1}{c}{\textcolor{blue}{$\uparrow1.62\%$}} &
  \multicolumn{1}{c}{\textcolor{blue}{$\uparrow0.28\%$}} \\ \bottomrule
\end{tabular}}
\label{tab:memotion-tasks}
\end{table*}

\subsection{Evaluation on the \memotion\ dataset}
\label{sec:memotion}
We then compare the performance of \proposed, other baselines, and the state-of-the-art systems on the \memotion\ shared task \cite{sharma2020semeval2020}. 
The \memotion\ dataset contains approximately $8K$ memes. It was proposed for the  three subtasks\footnote{We use abbreviations \sent,\ \emot\ and \emotq\ for \textit{sentiment analysis}, \textit{emotion classification}, and \textit{emotion class quantification}, respectively.} -- \textit{sentiment analysis} (positive/negative), \textit{emotion classification} (humour/sarcasm/offense/motivational), and \textit{emotion class quantification} (slightly/mildly/very).
The original average baseline F1 for the three \memotion\ sub-tasks -- \sent\ ($0.2176$), \emot\ ($0.5002$), and \emotq\ ($0.3009$) -- 
indicate inherent non-triviality of the tasks. The previous best systems involve a word2vec \cite{NIPS2013_9aa42b31,Mikolov2013EfficientEO} based feed-forward neural network for \sent\ \cite{keswani-etal-2020-iitk-semeval}, a multimodal multi-tasking based setup for \emot\ \cite{vlad-etal-2020-upb}, and a feature-based ensembling approach for the \emotq\ task \cite{guo-etal-2020-guoym}. The performance of the state-of-the-art systems for the three tasks (c.f. Table \ref{tab:memotion-tasks}) are $0.3547$, $0.5183$, and $0.3225$ F1, respectively. In comparison, \proposed\ induces an increase of $0.89\%$ and $0.28\%$ F1-score for \emot\ and \emotq\ tasks, respectively. For \sent, however,  \proposed\ lags slightly behind the \textit{previous best} score by 0.61\% F1 score.

\proposed's\ low score on the \sent\ task could be due to noise induced by the emotion-enriched feature that might complicate modeling a more straightforward task like \sent\ compared to simpler early-fusion-based state-of-the-art.
\proposed\ also performs relatively better with $1.09\%-1.13\%$ F1-score increment in the \textit{humour} and \textit{sarcasm} categories for the \emot\ task, as against that for \emotq. Since the level of abstraction for the information being modeled for \emot\ (emotion classification) is relatively higher as compared to that for \emotq\ (emotion quantification), an explicit emotion modeling could help detect emotions for \emot, and not necessarily for fine-grained emotion intensity quantification in \emotq, especially for complex categories like \textit{humor} and \textit{sarcasm}. On average, \proposed's\ performance on \memotion\ tasks are comparable -- it reports better scores for the \emot\ and \emotq\ tasks; however, it yields inferior performance in the \sent\ task.

\subsection{Ablation Studies}
\begin{wraptable}{r}{4.8cm}
    \caption{Ablation study for \proposed\ on \dataset; emotion features (EMO), GMF and GCA (via DCA: Dense Co-attention) modules.}
    \centering
    \resizebox{0.25\textwidth}{!}{
    \begin{tabular}{lcc}
    \toprule
    \multicolumn{1}{c}{\bf Approach} & \bf F1 & \bf Acc \\ \midrule
    \rowcolor{LightGreen}
    \proposed\ & 82.43 & 82.39 \\
    \quad -- EMO  & 77.19 & 77.08 \\ 
    \quad -- GMF  & 80.60  & 80.38 \\
    \quad -- GCA + DCA & 80.04 & 79.97 \\  \bottomrule
    \end{tabular}}
    \label{tab:ablation}
\end{wraptable}
The incorporation of emotion features induces 5\% improvement over \proposed\ without emotion features. Further, replacing the GMF module with simple concatenation for incorporating emotion features causes a 2\% performance drop. On the other hand, GCA is also observed to be pivotal as its replacement with dense co-attention (DCA) \cite{dense_co_attn} induces a drop of 2\% performance. Effectively, the exclusion of GMF and GCA from \proposed\ is empirically observed to induce a performance drop of $\approx 2\%$, as shown in Table \ref{tab:ablation}.

We analyze the contribution of each module of \proposed\ on \dataset: {emotion encoder}, gated multimodal fusion (GMF), and gated cross attention (GCA) in Table. \ref{tab:ablation}. 
\proposed\ without EMO is expected to perform worse due to the complexity of the model not being complemented by the required rich features. This corroborates the requirement of a solution that explicitly incorporates emotion-enriched feature modeling. The pre-training of the emotion encoder is discussed in detail in Section \ref{subsec:emopretrain}. On the other hand, the early fusion model, being efficient yet straightforward towards multimodal classification, yields impressive results, which we also consider the best baseline for comparison. Also, besides discussing the effect of \proposed\ without EMO (c.f. Table \ref{tab:ablation}), we specifically include evaluations without GMF and GCA to examine the optimal modeling of emotion features using these modules in \proposed. For both exclusions, the performance is low. This assessment consolidates the boosting capacity of each module constituting \proposed.

\subsection{Discussion}
\noindent \textbf{-- Novelty Aspects of \proposed:}
We empirically establish the efficacy of using low-rank bilinear pooling-based non-linear gating fusion instead of simple concatenation for fusing emotion-aware image representations in GMF. This helps characterize intra-modal fusion against inter-modal fusion, for which concatenation has been the conventional fusion strategy. For the GCA module, we first perform the conditioning based upon the emotion-aware \textit{image} representation to compute the \textit{textual} attention and feed it back towards computing a final emotion-aware image representation. This is in contrast to the convention of performing either a text-based conditioning \cite{adaptive_co_attn} or a parallel co-attention based strategy \cite{lu2019vilbert}. To our understanding, this is the first attempt to emphasize the visual cues toward overall modeling.  
    
Essentially, through our proposed approach incorporating the adaptation of existing effective techniques like GMF and GCA, we present a strategy that has been observed to be empirically adequate for modeling intra-modal and inter-modal fusion. To the best of our understanding, incorporating emotion-oriented features explicitly via \textit{visual} modality toward a task like a meme analysis has not been explored prior to this work.

\noindent \textbf{-- Dataset Utility:}
Meme datasets mostly  encapsulate affective dimensions representing higher levels of abstraction ranging from categories like humor, sarcasm, offense, and motivation in the case of Memotion, to aspects like Harmfulness and Hatespeech within memes in \harmeme\ and \dank\ datasets, respectively. Additionally, since memes in such datasets usually capture real-world events involving famous personalities and phenomena, they tend to be reasonably restricted in terms of the visual subjects they embody. For instance, a significant portion of such memes does not contribute visually towards affective adjudication, limiting the characterization due to cartoons, caricatures, and expressive personifications, for meme emotion detection. Most of them typically end up projecting textual cues as their characteristic feature.
    
In contrast, \dataset\ captures the affective dimension that objectively focuses on six \textit{basic} \textit{Ekman} emotions via multimodal cues for \textit{generic} themes, capturing the expressivity differently from other datasets and soliciting an appropriate investigative framework. Such memetic configuration aptly represents the scope of this work -- detecting basic emotions from \textit{generic memes}, which tend to bear emotional expressivity via both image and text modalities.

\begin{figure*}[t!]
    \centering
    \includegraphics[width=0.85\textwidth]{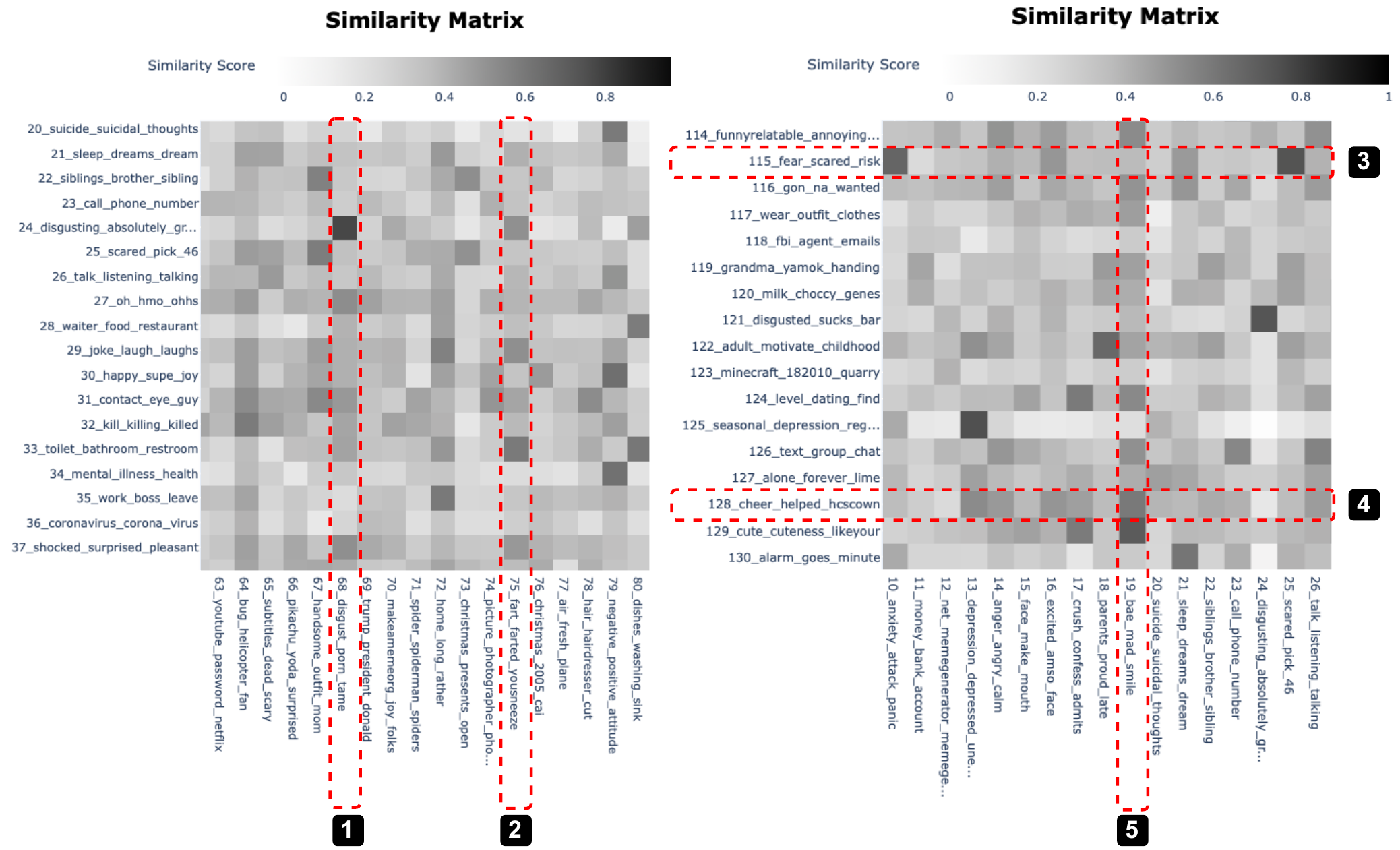}
    \caption{Thematic overlap analysis via similarity matrix, with \textit{five} example cases of inter-emotion overlap highlighted. Each x/y-axes label represents \textit{topic id}, followed by a set of \textit{three} corresponding \textit{representative key-words} (\texttt{topicid\_kw1\_kw2\_kw3}).}
    \label{fig:sim}
\end{figure*}

\begin{figure}[t!]
    \centering
    \includegraphics[width=0.9\columnwidth]{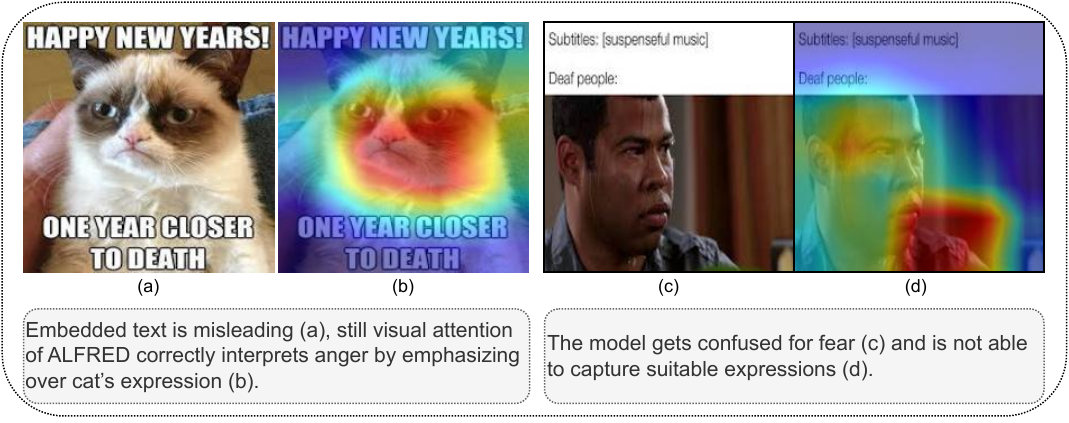}
    \caption{Depiction of: Interpretability Analysis [subfigs. (a) and (b) -- \proposed\ correctly predicts an \textit{anger} meme], \texttt{Error Analysis} [subfigs. (c) and (d) -- \proposed\ incorrectly predicts a \textit{fear} meme as a \textit{surprise} meme].}
    \label{fig:int_err}
\end{figure}

\subsection{Interpretability}

We attempt to interpret the decisions made by \proposed\ using GradCam \cite{gradcam}. It uses gradients flowing through a model to produce a rough attention map. This highlights the regions in the image that the model pays attention to while making a decision.

In Fig. \ref{fig:int_err}(a), the text alone is not sufficient to detect the emotion of the meme, as the excerpt `\texttt{HAPPY NEW YEARS!...}' could mislead the model's decision. The key aspect here lies with the \textit{angry} (`grumpy cat meme') expression of the cat. In Fig. \ref{fig:int_err}(b), we notice that the model pays attention to the cat's face to correctly predict that the emotion is \textit{anger}.

\subsection{Error Analysis}
\noindent \textbf{-- Visual Obscurity:} On analyzing the incorrect predictions from the test set, we find that most misclassifications involve complex memetic text or obscure visuals, including prominent visual occlusion. Another distinct trend for poor results is observed for the samples belonging to the \textit{least represented} emotion categories \textit{fear} and \textit{disgust}, with 2.5 $\%$ decrease and a marginal enhancement of 0.62 $\%$ respectively, in the accuracy values compared with those from an \textit{early-fusion} based model. An example that demonstrates the misclassification attempt of \proposed\ due to both the aforementioned likely reasons is shown in Fig. \ref{fig:int_err}(c), depicting the facial expressions of the man to be that of \textit{fear}. Still, the model incorrectly predicts it as \textit{surprise}. On interpreting the visual attention-map via GradCAM-based visualization, we observe that \proposed\ does not pay attention to the subtle facial expressions indicating \textit{fear}, as demonstrated by the misplaced visual attention, in Fig. \ref{fig:int_err}(d). 

\noindent \textbf{-- Textual Obscurity:} Data sufficiency is also observed to play pivotal role towards class-wise performances observed in Table \ref{tab:task-0-combined} w.r.t. \textit{textual influence} within memes. A critical evidence corroborating this aspect is the lexical richness of memetic text for categories \textit{disgust} and \textit{fear}, as against lexical obscurity for \textit{surprise}, as observed from Figs. \ref{fig:wordclouds}(f), \ref{fig:wordclouds}(a) and \ref{fig:wordclouds}(e), respectively. The former two, being sparsely represented within \dataset, yield sub-par performances for the corresponding categories, whereas the latter being densely represented, contributes a decent accuracy (c.f. Table \ref{tab:task-0-combined}), despite lexical obscurity. This suggests that multimodal (as against unimodal) contextual dependency is imperative towards emotion recognition from memes since it is not just the complex cross-modal interplay that encapsulates the intended message but the modality-specific intricacies that constitute complex memetic designs.

\noindent \textbf{-- Analyzing Thematic Overlaps:} We further investigate the semantic complexity posed by the themes that various memes are based on. To this end, thematic structure and characteristics are derived via a clustering-based approach and are ascertained for semantic overlap \cite{grootendorst2022bertopic} w.r.t. the six \textit{Ekman} emotions for \dataset. Firstly, document embeddings are obtained via \texttt{all-MiniLM-L6-v2} based Transformer model \cite{reimers-gurevych-2019-sentence}, followed UMAP-based dimensionality reduction \cite{McInnes2018} and HDBSCAN-based clustering \cite{hdbscan}.

The hierarchical thematic sub-groupings obtained through HDBSCAN, analyzed for semantic similarity using similarity matrices (shown in Fig. \ref{fig:sim}), reveal distinct overlaps and proximity regarding six of Ekman's emotions. These patterns are illustrated through highlighted examples. Notably, \textit{disgust} memes (topic id: $68, 75$) represented by patterns $\#1$ and $\#2$ in Fig. \ref{fig:sim} show variable overlaps with \textit{surprise} (topic id: $37$) and \textit{joy} (topic id: $29, 30$).

The second similarity matrix in Fig. \ref{fig:sim} highlights patterns connecting \textit{fear} (topic id: $115$) with \textit{anger} (topic id: $14$) and \textit{joy} (topic id: $16$) (pattern $\#3$), and \textit{joy} (topic id: $128$) with \textit{sadness} (topic id: $13$) and \textit{anger} (topic id: $14$) (pattern $\#4$). Additionally, \textit{anger} (topic id: $19$) overlaps with \textit{sadness} (topic id: $127$) and others, along with \textit{joy} (topic id: $128, 129$) in pattern $\#5$.

Overall, \textit{joy} and \textit{sadness} consistently emerge as common factors in emotion overlaps, aligning with Ekman's \cite{basic_emotions_ekman} and Plutchik's \cite{Plutchik1988} theories. This suggests that the proximity of these emotions in memes stems from complex linguistic content. Determining the exact valence of this content remains challenging, underlining the need for detailed emotion analysis of memes.

\subsection{Generalizability}
\label{sec:general}
Here, we establish the generalizability of \proposed\ for \harmeme\ \cite{harmeme} and \dank\ tasks \cite{dank_evalita}. The dataset for \harmeme\  constitutes $\approx7K$ memes (in English) on Covid-19 and US Politics. This dataset captures annotations for harmfulness and the targeted entity types. The second dataset, \dank, comprises $\approx1K$ hateful memes (in Italian). The memes are about the 2019 Italian Government Crisis. There was an associated shared task involving three subtasks -- a) meme detection, b) hate-speech identification, and c) event clustering. In this work, we focus on hate-speech identification to ensure evaluation consistency.   

\begin{figure}[!t]
    \centering
     \includegraphics[width=0.9\columnwidth]{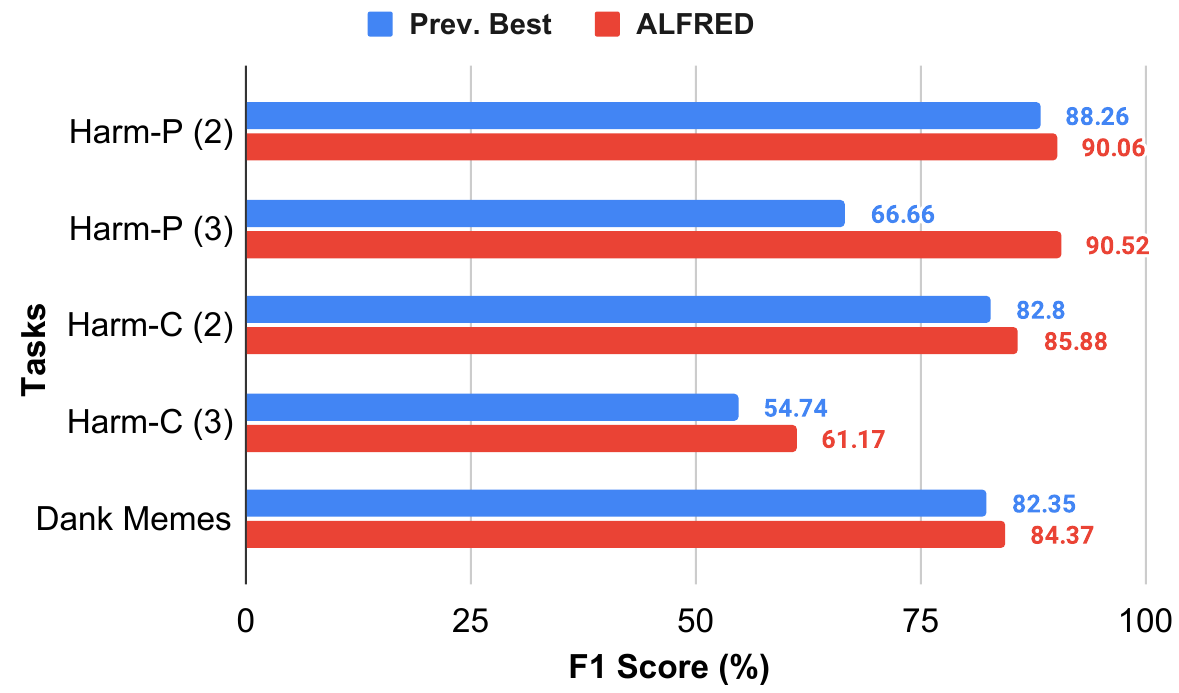}
     \caption{Performance comparison for \proposed\ and \textit{previous best} on \harmeme\ (US Politics (P)/Covid (C); 2/3 class classification) and \dank\ tasks, demonstrating \proposed's\ generalizability.}
     \label{fig:f1_score_gen}
\end{figure}

\noindent \textbf{-- \harmeme:\ } The best performance on this dataset was reported by MOMENTA \cite{pramanick-etal-2021-momenta-multimodal} which strongly outperformed sophisticated multimodal baselines such as V-BERT and ViLBERT. For two-class classification, \proposed\ is observed to achieve an improvement of 3.08\% and 1.8\% F1 over MOMENTA, respectively, on the Harm-C and Harm-P datasets. For  three-class classification, \proposed\ achieves 6.43\% and 23.86\% F1 increment over MOMENTA on the Harm-C and Harm-P datasets, respectively (c.f. Fig. \ref{fig:f1_score_gen}). 

\noindent \textbf{-- \dank:\ }Dank Memes is an Italian hateful politics meme dataset. The top two submissions for the related shared-task were both early-fusion based: \textit{Unitor} employs domain-specific pretraining 
before finetuning on Dank Memes; \textit{UPB} uses VGCN-BERT for text modality \cite{dank_evalita}. Since the task deals with memes with embedded Italian content, we replace the BERT model with UmBERTo \cite{umBERTo} within \proposed\ while keeping other components same. \proposed\ achieves an absolute increment of 2.02\% and 4.37\% in F1 and precision, respectively, over the best baseline, while the recall lags by 1.96\% behind \textit{Unitor}. These results not only highlight \proposed's\ generalizability, but also indicate its language-agnostic cross-lingual affinity (c.f. Fig. \ref{fig:f1_score_gen}), especially w.r.t multimodal tasks like meme analysis.

\section{Conclusion}

In this work, we first introduced \dataset,\ a new dataset for detecting emotions in Internet memes. We then proposed \proposed, which uses emotion-aware meme representations to detect emotions from memes. Extensive experiments indicated that \proposed\ outperforms strong multimodal baseline with 4.94\% F1 increment and yields robust performance on the \memotion\ task \cite{sharma2020semeval2020} dataset. Further, we investigated the interpretability of the model by establishing the correspondences between the correct emotion class being predicted and the expressive emotions being attended to within the meme image. We also highlighted the inherent limitations that explicit emotion modeling can develop. Finally, we established the generalizability of \proposed\ by demonstrating its superiority over \textit{previous best} baselines on the \harmeme\ and \dank\ datasets. As part of the future extension to this work, we would like to explore a multi-task learning setup involving the detection of correlated fine and coarse-grained emotion features for memes. Moreover, tasks like explanation generation for various meme-emotions and network structure-based investigation of meme virality, w.r.t the emotions, are also promising avenues to explore. 


%

\section*{Acknowledgment}
T. Chakraborty would like to acknowledge the support of the Wipro Research Grant.

\ifCLASSOPTIONcaptionsoff
  \newpage
\fi



\bibliographystyle{IEEEtran}
\bibliography{custom}
%



%
\vspace{-1.2cm}
\begin{IEEEbiography}[{\includegraphics[width=2cm,keepaspectratio]{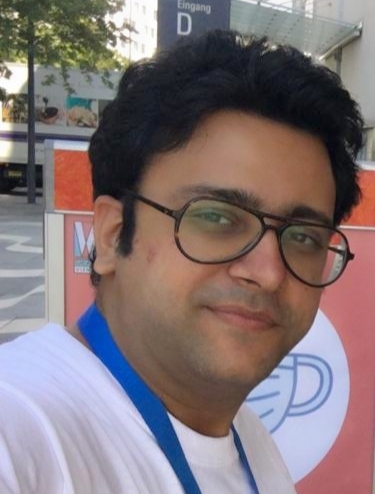}}]{Shivam Sharma}
is a Ph.D. student in the Dept. of Electrical Engineering at Indian Institute of Technology Delhi (IIT Delhi), India. His research interests span around multimodal applications within NLP. He also works as a Lead Research at Wipro AI Research (Lab45), Wipro Ltd. He completed his MS (by Research) in Signal Processing (Acoustics) from the Indian Institute of Information Technology, Sri City (IIIT-S) in 2020.
\end{IEEEbiography}\vspace{-1.4cm}
\begin{IEEEbiography}[{\includegraphics[width=1.8cm,keepaspectratio]{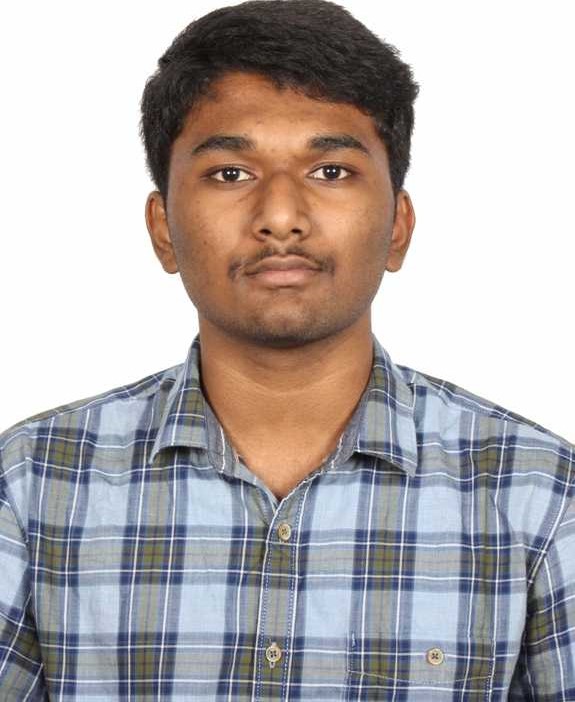}}]{Ramaneswaran S}
works as a Data Scientist at NVIDIA. He completed his Bachelor of Information Technology in the School of Information Technology and Engineering at Vellore Institute of Technology (VIT), Vellore, Tamil Nadu, India. His broad research interests include Natural Language Processing and Conversational AI.
\end{IEEEbiography}\vspace{-1.9cm}
\begin{IEEEbiography}[{\includegraphics[width=2.1cm,keepaspectratio]{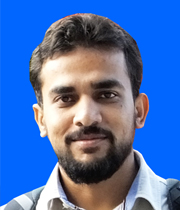}}]{Md. Shad Akhtar} 
is an Assistant Professor at Indraprastha Institute of Information Technology, Delhi (IIIT-D). He completed his Ph.D. in Computer Science and Engineering from IIT Patna in 2019. He received his M.Tech from IIT (ISM), Dhanbad. He also has over two years of industry experience with HCL Tech. Ltd. His main research areas are Sentiment and Emotion Analysis in the Natural Language Processing domain. Currently, his area of interest focuses on Dialog Management and Multimodal Analysis.
\end{IEEEbiography}\vspace{-1.1cm}
\begin{IEEEbiography}[{\includegraphics[width=2.3cm,keepaspectratio]{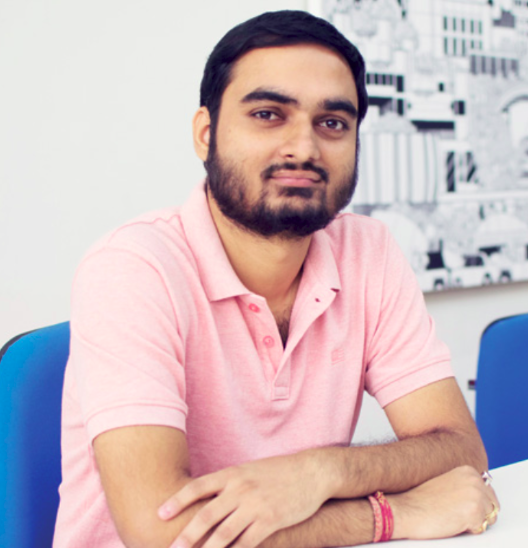}}]{Tanmoy Chakraborty}
is an Associate Professor of Electrical Engineering and an Associate Faculty member of the Yardi School of AI at IIT Delhi. He leads the Laboratory for Computational Social Systems (LCS2), a research group specializing in NLP and Computational Social Science. His current research primarily focuses on empowering frugal language models for improved reasoning, grounding, and prompting. He received numerous awards, including the Ramanujan Fellowship, PAKDD Early Career Award, ACL'23 Outstanding Paper Award, IJCAI'23 AI for Good Award, and several faculty awards/gifts from industries like Facebook, Google, LinkedIn, JP Morgan, and Adobe. He has authored a textbook on "Social Network Analysis". More details may be found at tanmoychak.com.

\end{IEEEbiography}
\clearpage

\appendices

\section{Additional details of \dataset}
This section provides additional details on collecting and curating our proposed dataset \dataset.

\subsection{Filtering Criteria}
\label{app:subsec:data}
For downloading meme images, we used the Mozilla Firefox extension tool, called Download All Images\footnote{\href{https://addons.mozilla.org/en-US/firefox/addon/save-all-images-webextension}{Firefox Browser ADD-ONS | Download All Images} \ExternalLink}, with a few downloading specifications configured. These were file size (min): 4 KB, dimensions: 200X200, and format: JPGs and PNGs. We set these specifications after carefully observing the sample quality of memes available online and the requirements of the task at hand. Since despite pre-setting the required specifications, the download process ended up collecting images that were still unsuited towards manual annotations, the annotators were asked to further manually filter out images based on \textit{filtering criteria} specified in Section \ref{subsec:data} in the main text. These factors involved inadequate image resolution and text readability (perceptually ambiguity), absence of any of the six Ekman emotions, harmful memes containing personal information, and memes containing non-English textual content. Our primary heuristic for keeping a meme was the perceived intelligibility w.r.t. the textual and visual cues present in it. This ensured better interpretability of the model outputs as well. We did not consider any pre-defined (or otherwise) resolution threshold after collecting the raw meme images.

\subsection{Data Imbalance}
\label{app:subsec:imbalance}
Here, we want to highlight a popular effort towards investigating hateful memes via Hateful Memes Challenge \cite{kiela2020hateful}. This, although involved the curation of a balanced combination of hateful and non-hateful memes focusing on modality-specific nuances, did involve the inclusion of benign confounders towards evaluating the robustness of multimodal systems, but were created synthetically by adopting confounding strategies, essentially not reflecting the realistic data distribution. This effect has been empirically observed to exacerbate when evaluated for the content over other social media platforms like 4chan (/pol/) \cite{jennifer2022feels, Zannettou2018}. Keeping in mind the adverse implication of the non-realistic dataset, we instead emphasized collecting and curating a dataset that not only captures the fine-grained aspects of the primary task we intended to address but also reflects the realistic distribution, offering the scope for imminent developments and hence novelties in the areas like un/self-supervised and few-shot learning. This has already demonstrated  capabilities for characterizing harmful content over social media platforms\footnote{\href{https://ai.meta.com/blog/harmful-content-can-evolve-quickly-our-new-ai-system-adapts-to-tackle-it/}{Meta AI | ML Applications \ExternalLink}}.

\begin{table*}[t!]
\centering
\caption{Top 10 most frequent words in each emotion class. The TF-IDF score is in the parenthesis.}
\resizebox{0.8\textwidth}{!}{%
\begin{tabular}{c|c|c|c|c|c}
\hline
{\bf Fear}    & {\bf Anger} & {\bf Joy}      & {\bf Sadness}    & {\bf Surprise} & {\bf Disgust} \\ \hline
mom (27.2419)    & face (40.5642)   & love (56.7116)    & depression (45.3137) & realize (17.5789) & absolutely (38.0648) \\ \hline
scared (21.0282) & mad (33.5129)  & day (38.6249)     & life (40.7641)      & like (16.5175)    & face (11.9207)   \\ \hline
pick (15.2956)   & like (29.2126) & excited (38.2704) & like (38.8570)      & oh (14.6915)      & people (11.6430) \\ \hline
people (13.5623) & know (23.5728) & friend (36.5688)  & anxiety (33.9166)   & meme (13.0155)    & make (9.8716)    \\ \hline
hear (12.0607)   & make (23.2726) & mom (35.4429)     & day (31.8213)       & people (12.3859)  & food (7.7485)    \\ \hline
spider (11.9407) & just (22.9047) & good (35.2838)    & depressed (29.3055) & time (12.2300)    & meme (7.3623)    \\ \hline
afraid (11.3161) & people (22.6959) & friends (33.2909) & lonely (29.1180)     & just (10.7769)    & look (6.8671)        \\ \hline
says (10.9221)   & look (21.8237) & like (27.5951)    & going (28.7829)     & mom (10.2412)     & like (6.3189)    \\ \hline
home (9.0328)    & time (20.6286) & make (26.5385)    & friends (26.8488)   & face (9.2913)     & realize (6.1869) \\ \hline
time (8.6592)    & say (20.5275)  & best (25.1675)    & feel (26.7687)      & hell (8.8602)     & just (6.0652)    \\ \hline
\end{tabular}
}
\label{tab:top-words}
\end{table*}
\subsection{Thematic Analysis}
\label{app:subsec:themes}

Towards performing thematic analysis for \dataset, we leverage a popular topic modelling technique, called BERTopic \cite{grootendorst2022bertopic} that uses transformers and c-TF-IDF to create dense clusters. For the thematic analysis of visual objects, the overall pipeline first converts images into embeddings, followed by the performing dimensionality reduction, followed by HDBSCAN-based dense clustering. This is followed by captioning the images while weighting the cluster representative bag-of-words using c-TF-IDF and finding the best matching images based on most representative documents. Additionally, we also assess the tf-idf ranked set of words from each categorical distribution in \dataset, as shown in Table \ref{tab:top-words}.

\begin{figure}[th!]
    \centering
    \includegraphics[width=\columnwidth]{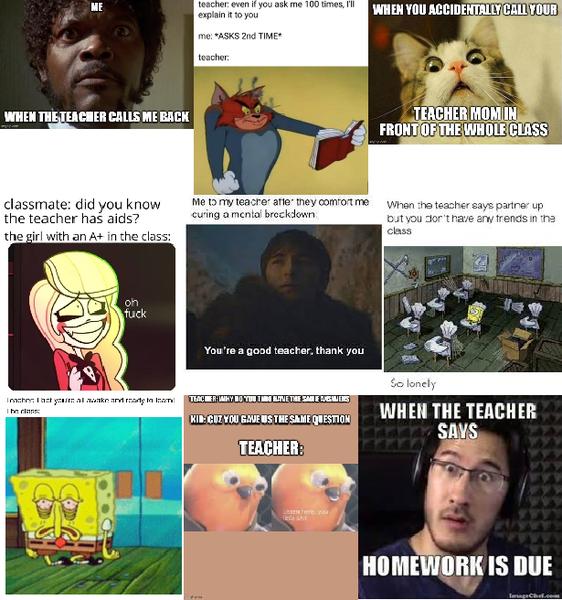}
    \caption{A collection of meme examples featuring \textit{human subjects, pop culture references, animated characters, and animals} from \dataset, with representative topics as \textit{teacher, class, school, homework, test, kid, and exam}.}
    \label{fig:visual_subjects}
\end{figure}

\begin{figure*}[t!]
    \centering
    \subfigure[\proposed\ (w. Frozen Emotion Encoder)]{\includegraphics[width=0.30\textwidth]{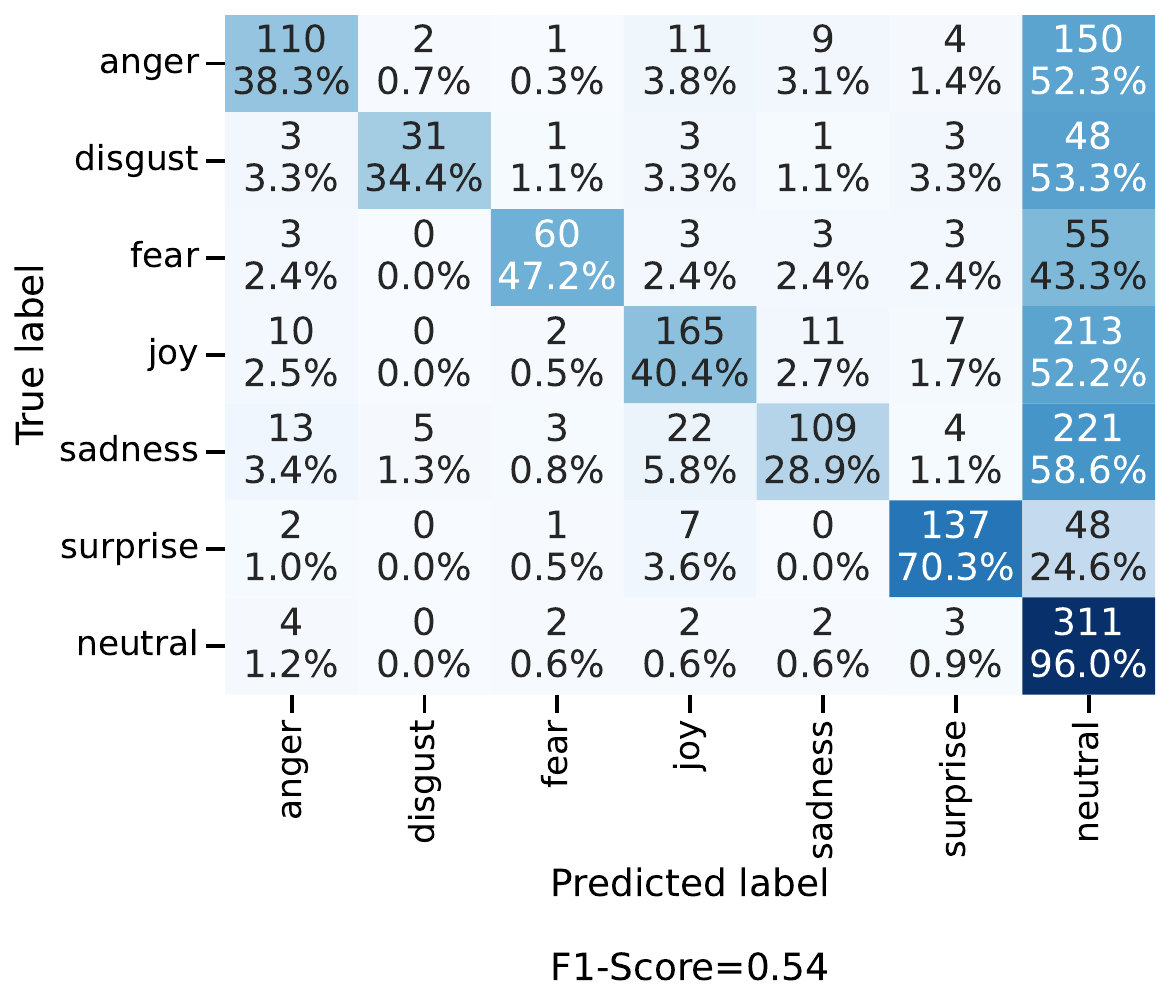}}
    \subfigure[\proposed\ (w. Fine-tuned Emotion Encoder)]{\includegraphics[width=0.30\textwidth]{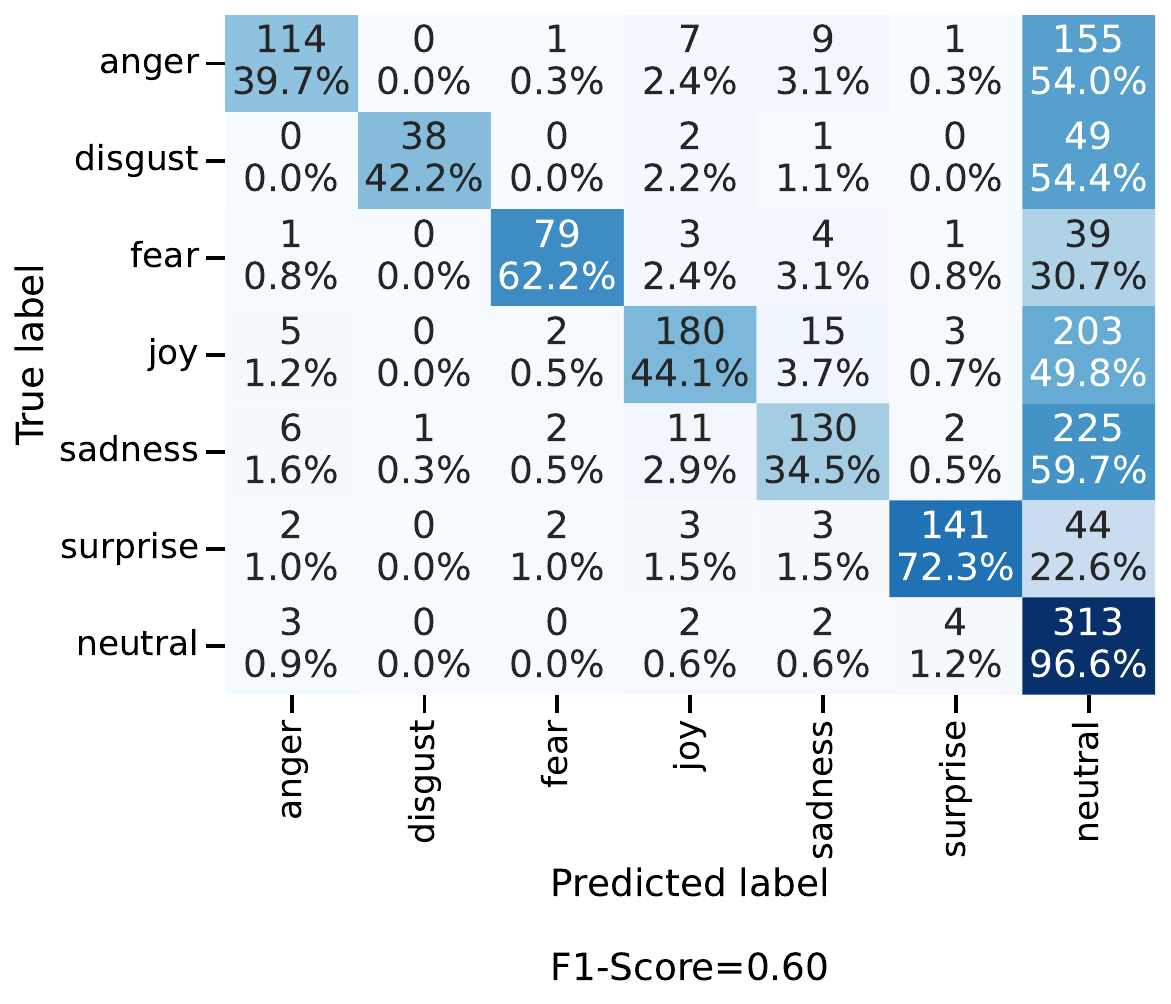}} 
    \subfigure[Category-wise Confusion ($\text{FN}_{\text{E}_{i}\rightarrow \text{N}}-\text{TP}_{\text{E}_{i}\rightarrow \text{E}_{i}}$)]{\raisebox{0.1\height}{\includegraphics[width=0.38\textwidth]{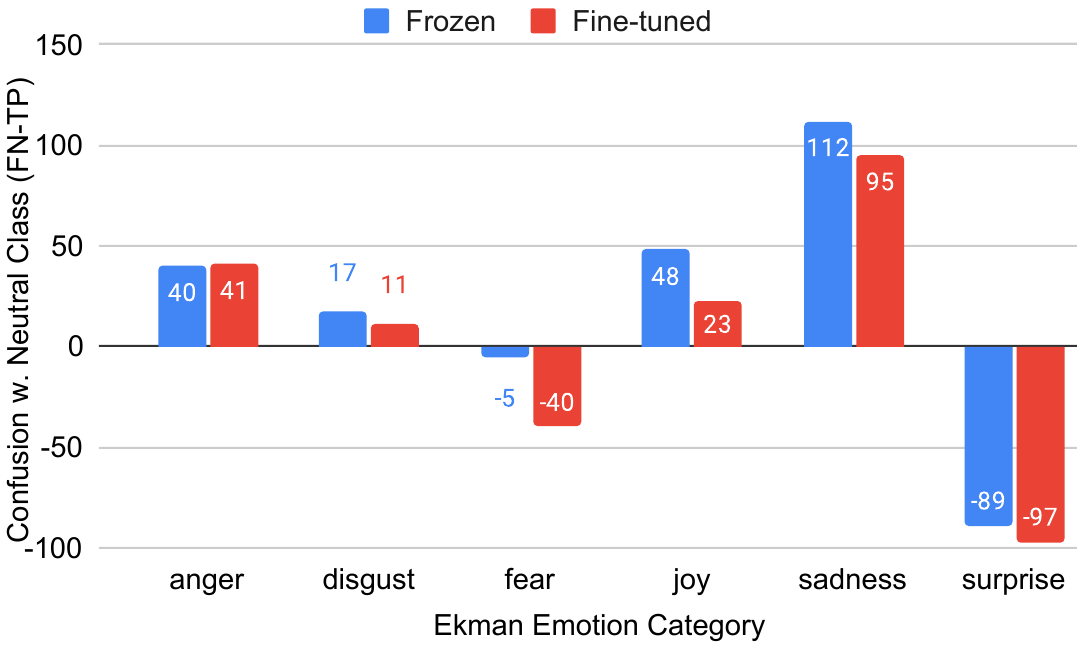}}}
    \caption{Analyzing \proposed's performance \textit{with} \textit{neutral} category. (a) and (b) Confusion Matrices and F1-score for \proposed's two variants; (c) Quantifying Confusion: $\text{FN}_{\text{E}_{i}\rightarrow \text{N}}-\text{TP}_{\text{E}_{i}\rightarrow \text{E}_{i}}$: Difference between the \textit{false-negatives} ($\text{FN}$) w.r.t \textit{neutral} class ($\text{N}$) and \textit{true-positives} ($\text{TP}$) for each Ekman emotion category ($\text{E}_{i}$).}
    \label{fig:neutres}
\end{figure*}


\begin{figure*}[th!]
    \centering
    \includegraphics[width=\textwidth]{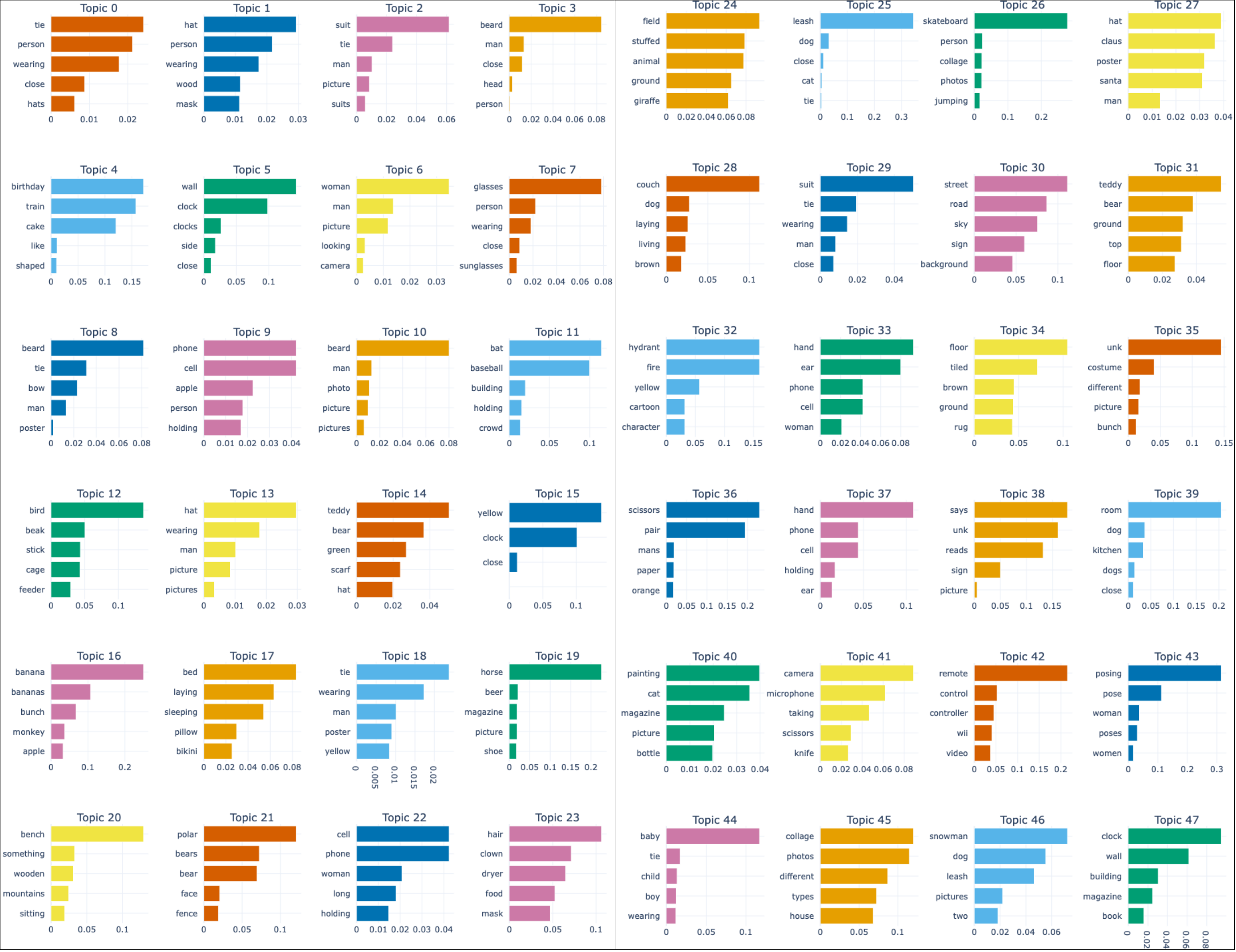}
    \caption{Top$-48$ prominent topics representing themes of the \textit{visually depicted content} in \dataset's memes.}
    \label{fig:captions_bar}
\end{figure*}

We look for the visual diversity captured within the MOOD dataset via manual and automated assessment. The manual review suggests a pre-dominant visual representation of human subjects, pop culture references, animated characters, and animals in the memes. In addition, the memes typically consist of various artistic modifications of these basic elements --\textit{visual morphing} and \textit{juxtapositioning}, along with diversified \textit{textual overlays}. A representative set of such samples from \dataset\ is shown in Fig. \ref{fig:visual_subjects} for \textit{third largest topic cluster} associated with meme's visual embeddings. This topic is defined by words \textit{teacher, class, school, homework, test, kid, and exam}. We choose this set for exemplification of \dataset's visual diversity, as it has a relatively more diverse set of meme templates and visual subjects utilized as variants, in comparison to that within the larger topic clusters defined by words - \textit{disgusting, happy, she, crush, girlfriend, cute}, which are dominated by template-based meme designs.
Further, we also analyze the bar charts of the top 48 topic clusters defined by the caption keywords corresponding to the image-embedding-based clusters. These are shown in Fig. \ref{fig:captions_bar}.

\begin{figure*}[th!]
    \centering
    \includegraphics[width=\textwidth]{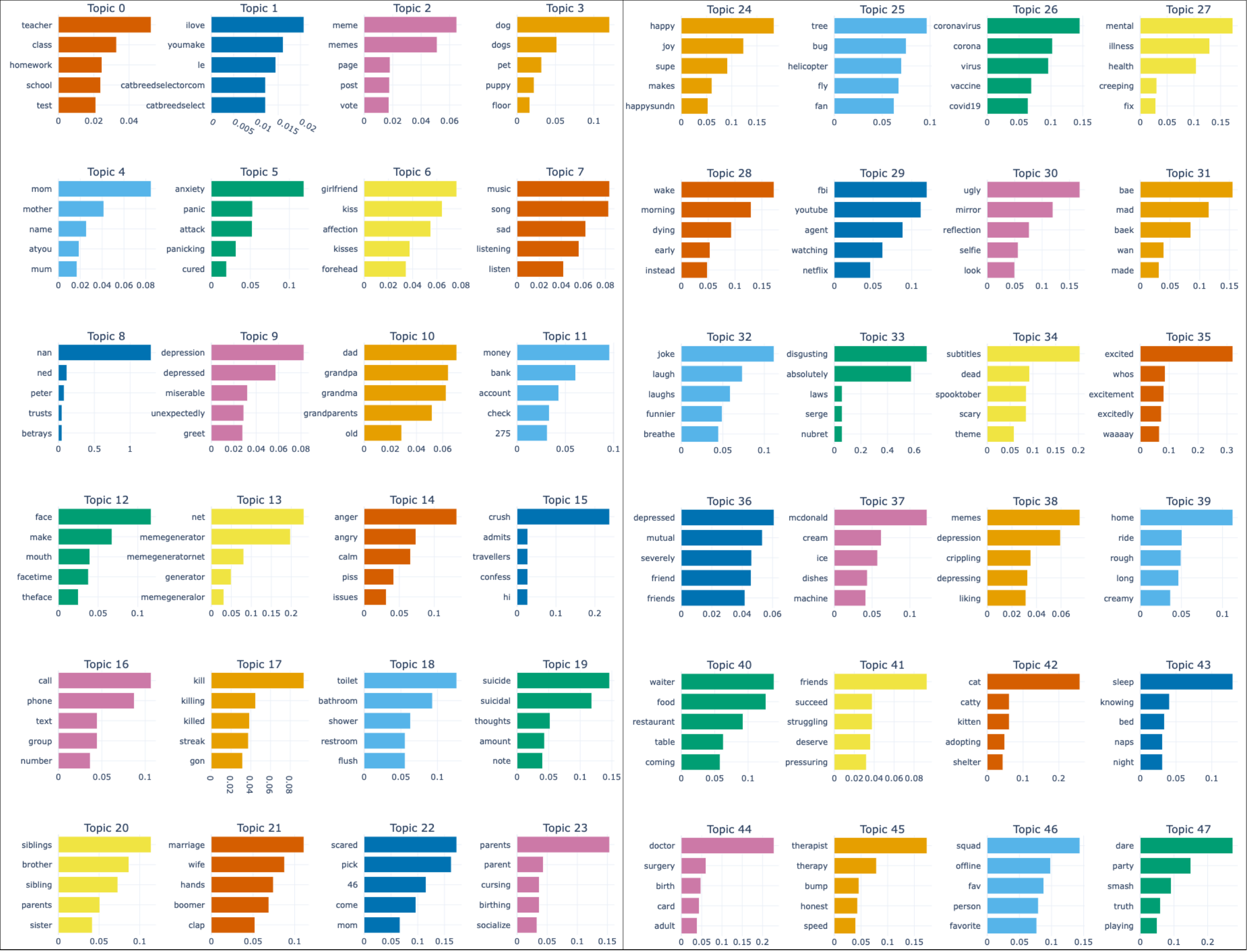}
    \caption{Top$-48$ prominent topics representing themes of the \textit{textually embedded content (OCR)} in \dataset's memes.}
    \label{fig:captions_bar}
\end{figure*}

\section{\proposed's Performance with `Neutral' Memes}
\label{apps:sec:neutdiscuss}

We examine \proposed's efficacy/constraints towards the multi-class classification setup involving \textit{fear, anger, joy, sadness, surprise,} and \textit{disgust}, along with a seventh category, `neutral'. We evaluate \proposed, by training it using memes from \textit{neutral} category as well. To this end, we utilize a total of $2197$ \textit{neutral} category memes ($1549$ memes in training, $324$ memes in validations, and $324$ memes in testing) from the publicly available dataset, \memotion\ \cite{sharma2020semeval2020}, which (along with the other categories in the dataset) is systematically curated towards the designated categories. We do not use memes that we discard during our data collection process towards considering \textit{neutral} class, as it mostly consists of \textit{low-quality, noisy, or harmful} content, the generalizability towards which is accounted for in more realistic settings, as part of Section \ref{sec:general} (main text). We compare the performance of \proposed\ when the emotion encoder weights are: (a) \textit{frozen}, and (b) \textit{fine-tuned}. As can be observed from the F1-scores in Figs. \ref{fig:neutres} (a) and (b), the overall performance of \proposed\ gets reduced from $0.82$ F1-score, when modeled for only \textit{six} Ekman emotions, to the sub-par scores of $0.54$ and $0.60$ F1-scores for \textit{frozen} and \textit{fine-tuned} emotion-encoder-based scenarios, respectively. This highlights the limitations that just training on Ekman emotion-based samples, without considering the confounding effect of \textit{neutral} class, can get induced within \proposed's performance.

The confusion matrices for these experiments are also shown as part of Figs. \ref{fig:neutres}(a) and (b), respectively. The overall relative performance pattern in terms of the difference b/w Ekman emotion category-specific \textit{true-positives} (TP) and \textit{neutral} class (FN) is distinctly reflected in Fig. \ref{fig:neutres}(c). We observe that all the Ekman emotion categories get mixed up in different proportions with the \textit{neutral} category, with the \textit{most} confused class being \textit{sadness} with an FN-rate of $58.6\%$ and $59.7\%$ for \textit{frozen} and \textit{fine-tuned} variants of \proposed, respectively. At the same time, the \textit{least} confused category is \textit{surprise}, with an FN-rate of $24.6\%$ and $22.6\%$ for the corresponding variants. This observation, along with the slightly better accuracy produced by CLIP-based (reference) baseline ($0.764$) for class \textit{sadness} (c.f. Fig. \ref{tab:task-0-combined}) hints at the utility of leveraging more \textit{contextually enriched} representations towards discriminating it against the rest. Moreover, the distinct clarity of \proposed\ towards discriminating a class like \textit{surprise} is also corroborated by an imposing $6\%$ lead against our reference baseline (also having the second best category-specific) score. It is also worth noting that only minor confusions for \textit{neutral} class, being predicted as any other emotion category, are observed.

Additionally, the general trend of distinct reduction as shown in Fig. \ref{fig:neutres}(c), in \textit{differences} between the \textit{true-positive rate} (TPR) for Ekman emotions and \textit{false-negative rate} (FNR) w.r.t the \textit{neutral} class, when the emotion-encoder in \proposed\ is \textit{fine-tuned} (over the \textit{frozen} variant), clearly prescribes the effect of \textit{adapting} the emotion-encoder module, towards overall emotion classification. With the subtle exception of \textit{anger} class (exhibiting the enhancement of TPR-FNR difference by one sample), all the other classes project reasonable reductions in the overall confusion between Ekman emotions and \textit{neutral} category, quantified by the \textit{absolute} differences of $6, 35, 25, 17,$ and $8$ samples for classes: \textit{disgust, fear, joy, sadness,} and \textit{surprise}, respectively.

The amount of confusion visible between Ekman emotions and the neutral category suggests further scope of improvement in terms of out-of-distribution generalizability for \proposed.

\begin{figure}[t!]
    \centering
    \includegraphics[width=\columnwidth]{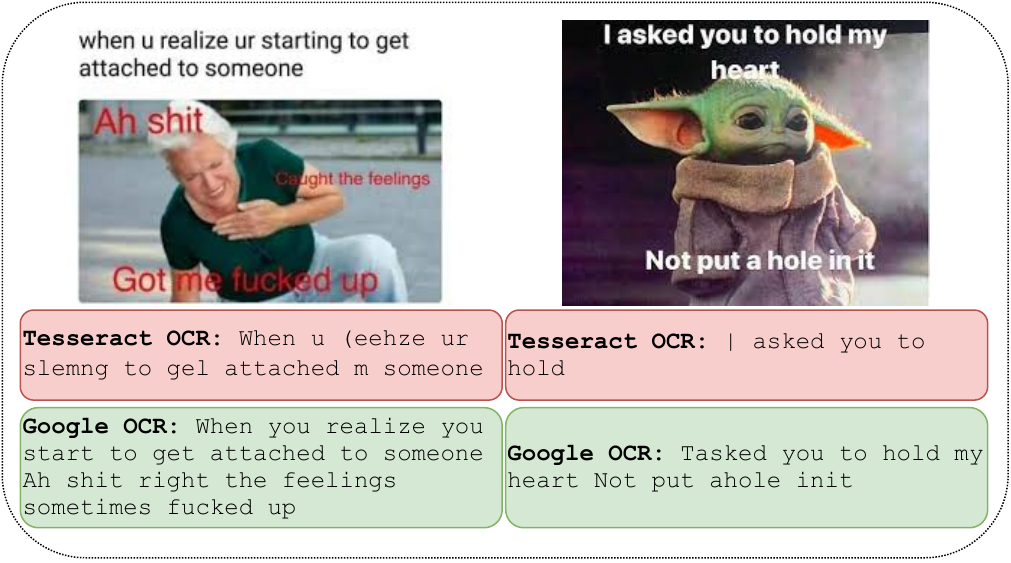}
    \caption{Comparison b/w the quality of the OCR-extracted text via (a) Tesseract OCR, and (b) Google OCR.}
    \label{fig:ocr_comp}
\end{figure}

\section{Text Extraction via OCR}
Text extraction via \textit{optical character recognition} (OCR) is of critical importance when mining embedded text from memes. The quality of the OCR process utilized influences the overall modeling capacity of systems. Towards exploring an optimal OCR technique for our purpose, we compare the text extractions for \textit{two} popular OCR-based text extraction APIs: Google Tesseract API\footnote{\href{https://pypi.org/project/pytesseract/}{Google's Tesseract-OCR API} \ExternalLink} (TOCR) and Google GCV API (GOCR). We first qualitatively analyze the extraction quality for 30 random memes and find occasional mistakes by TOCR, and rare by GOCR. For TOCR, mistakes committed were mostly for difficult cases, like text-image embedded at the same location, poor quality graphics, small text, etc. Sometimes even for simple cases, we observe GOCR's text quality much better than TOCR's output. A couple of examples shown in Fig. \ref{fig:ocr_comp} demonstrate the difference in the text-extraction quality for TOCR and GOCR. The first example shown in Fig. \ref{fig:ocr_comp} (\textit{left}) is the case consisting of a mix of simple and complex regions like black text on white background and ambiguous visual-text overlap, respectively, that cannot be correctly mined by TOCR, while GOCR, is distinctly more accurate in its extraction. On the other hand, the second, relatively simpler meme in Fig \ref{fig:ocr_comp} (\textit{right}) poses more obscurity to TOCR, as compared to better visibility for GOCR.

We also examine \proposed's overall performance in terms of the macro F1-score for our primary task of emotion classification for six Ekman emotions, w.r.t the two choices of OCR techniques explored. In drastic contrast to the impressive F1-score of $0.82$ observed for the GOCR-based text extraction, we find an abysmal show of performance by TOCR, with an F1-score of $0.75$, which speaks volumes of its inferiority when compared with Google GCV-based OCR API. In addition to leveraging GOCR for our primary experiments, we also conclude the critical influence that the correct OCR extraction technique has over the downstream task at hand.



\end{document}